\documentstyle[11pt]{article}
\normalbaselines

\newtheorem{thm}{Theorem}[section]
\newtheorem{lem}{Lemma}[section]
\newtheorem{prop}{Proposition}[section]

\begin{document}
\bibliographystyle{unsrt}

\def\bea*{\begin{eqnarray*}}
\def\eea*{\end{eqnarray*}}
\def\ba{\begin{array}}
\def\ea{\end{array}}
\count1=1
\def\be{\ifnum \count1=0 $$ \else \begin{equation}\fi}
\def\ee{\ifnum\count1=0 $$ \else \end{equation}\fi}
\def\ele(#1){\ifnum\count1=0 \eqno({\bf #1}) $$ \else \label{#1}\end{equation}\fi}
\def\req(#1){\ifnum\count1=0 {\bf #1}\else \ref{#1}\fi}
\def\bea(#1){\ifnum \count1=0   $$ \begin{array}{#1}
\else \begin{equation} \begin{array}{#1} \fi}
\def\eea{\ifnum \count1=0 \end{array} $$
\else  \end{array}\end{equation}\fi}
\def\elea(#1){\ifnum \count1=0 \end{array}\label{#1}\eqno({\bf #1}) $$
\else\end{array}\label{#1}\end{equation}\fi}
\def\cit(#1){
\ifnum\count1=0 {\bf #1} \cite{#1} \else 
\cite{#1}\fi}
\def\bibit(#1){\ifnum\count1=0 \bibitem{#1} [#1    ] \else \bibitem{#1}\fi}
\def\ds{\displaystyle}
\def\hb{\hfill\break}
\def\comment#1{\hb {***** {\em #1} *****}\hb }

\newcommand{\TZ}{\hbox{\bf T}}
\newcommand{\MZ}{\hbox{\bf M}}
\newcommand{\ZZ}{\hbox{\bf Z}}
\newcommand{\NZ}{\hbox{\bf N}}
\newcommand{\RZ}{\hbox{\bf R}}
\newcommand{\CZ}{\,\hbox{\bf C}}
\newcommand{\PZ}{\hbox{\bf P}}
\newcommand{\QZ}{\hbox{\rm eight}}
\newcommand{\HZ}{\hbox{\bf H}}
\newcommand{\EZ}{\hbox{\bf E}}
\newcommand{\GZ}{\,\hbox{\bf G}}

\font\germ=eufm10
\def\goth#1{\hbox{\germ #1}}
\vbox{\vspace{38mm}}

\begin{center}
{\LARGE \bf Fusion Operators in the Generalized $\tau^{(2)}$-model and Root-of-unity Symmetry of the XXZ Spin Chain of Higher Spin}\footnote{In this paper, the term ''XXZ spin chain'' means a lattice model with the $L$-operator associated to the trigonometric $R$-matrix of the ice-type model. The ''XXZ spin chain of higher spin'' here is named as the ''XXZ Heisenberg model with arbitrary spin'' in \cite{KiR}. }\\[10 mm] 
Shi-shyr Roan \\
{\it Institute of Mathematics \\
Academia Sinica \\  Taipei , Taiwan \\
(email: maroan@gate.sinica.edu.tw ) } \\[25mm]
\end{center}

\begin{abstract}
We construct the fusion operators in the generalized $\tau^{(2)}$-model using the fused $L$-operators, and verify the fusion relations with the truncation identity. The algebraic Bethe ansatz discussion is conducted on two special classes of $\tau^{(2)}$ which include the superintegrable chiral Potts model. We then perform the parallel discussion on the  XXZ spin chain at roots of unity, and  
demonstrate that the $sl_2$-loop-algebra symmetry exists for the root-of-unity XXZ spin chain with a higher spin, where the evaluation parameters for the symmetry algebra are identified by the explicit Fabricius-McCoy current for the Bethe states.  Parallels are also drawn to the comparison with the superintegrable chiral Potts model. 

\end{abstract}
\par \vspace{5mm} \noindent
{\rm 1999 PACS}:  05.50.+q, 02.20.Tw, 75.10Jm \par \noindent

\setcounter{section}{0}
\section{Introduction}
\setcounter{equation}{0}
In this paper, we investigate the uniform structure about symmetry of various lattice models\footnote{All the models discussed in this paper will always assume with the periodic condition. }, namely the generalized $\tau^{(2)}$-model, the chiral Potts model (CPM), and XXZ spin chain of higher spin. Parallels are also drawn to differentiating the symmetry nature appeared in those models.  The generalized $\tau^{(2)}$-model, also known as the Baxter-Bazhanov-Stroganov (BBS) model \cite{B89, B049, BazS, BBP, GIPS}, is the six-vertex model with a particular field (see, \cite{B049} page 3), i.e. having the $R$-matrix of the asymmetric six-vertex model, (see (\req(Rtau)) in this paper). Note that the non-symmetric diagonal Boltzmann weights of this twisted $R$-matrix differentiates it from the usual (gauge transformed) trigonometric $R$-matrix (see, for instance \cite{BIK} (3.3) and references therein), and the usual construction of $L$-operator by employing the highest-weight representation theory of quantum groups or algebras  fails in this context.  However using the cyclic $N$-vector representation of the finite Weyl algebra, one can construct a five-parameter family of $L$-operators for the $N$-state generalized $\tau^{(2)}$-model. When the parameters are restricted on a family of high genus curves, called the rapidities, one obtains the $\tau^{(2)}$-matrix of the solvable $N$-state CPM. In \cite{BazS} Bazhanov and Stroganov showed that the column-transfer-matrix of the $L$-operator in CPM possesses properties of the Baxter's $Q$-matrix (i.e., the chiral-Potts-transfer matrix) with symmetry operators similar to, but not exactly the same as, the $sl_2$-loop algebra.  Indeed  the $\tau^{(2)}$-degenerate eigenvalue spectrum occurs only in superintegrable case \cite{R04}, where the Onsager-algebra-symmetry operators in the quantum Hamiltonian chain \cite{GR} derived from the Baxter's $Q$-matrix provide the precise symmetry structure of the superintegrable $\tau^{(2)}$-model \cite{R05o}. Due to the lack of difference property of rapidities, the study of CPM, such as the calculation of eigenvalues \cite{B90, B93, MR} and order parameter \cite{B05}, relies on the method of functional relations among the $\tau^{(j)}$- and chiral-Potts-transfer matrices, consisting of (recursive and truncation) $T$-fusion relation, $TQ$- and $QQ$-relations. 
On the other hand, in the study of the "zero-field" six-vertex model (\cite{B04} Sect. 3) with the usual trigonometric $R$-matrix, it has been extensively analyzed in \cite{DFM, De05, FM01, R05b} that the degeneracy of the spin-$\frac{1}{2}$ XXZ Hamiltonian occurs when the anisotropic parameter $q$ is a root of unity with the extra $sl_2$-loop-algebra symmetry of the system \cite{DFM}. Recently, the $sl_2$-loop-algebra symmetry appears again in the XXZ chain of  spin-$\frac{N-1}{2}$ at the $N$th root of unity \cite{NiD}\footnote{The XXZ chain of  spin-$\frac{N-1}{2}$  at the $N$th root of unity in this paper is phrased as the Nilpotent Bazhanov-Stroganov model in \cite{NiD}.}. Results about the root-of-unity symmetry of the six-vertex model were first discovered by algebraic Bethe ansatz and quantum group theory,  later by the method of Baxter's $Q$-operator, and the degeneracy of eigenvalues was further found in the root-of-unity eight-vertex model  \cite{De01a, De01b, FM02}. By this effort, Fabricius and McCoy observed the similarity between the root-of-unity eight-vertex model and CPM, and proposed the conjectural functional relations of the eight-vertex model by encoding the root-of-unity symmetry property in a proper $Q$-operator of the theory, as an analogy to functional relations in CPM \cite{FM04}. Along this line, the $Q$-operator incorporated with the $sl_2$-loop-algebra symmetry of the spin-$\frac{1}{2}$ XXZ chain at the roots of unity was constructed in \cite{R06Q} where functional relations in the Fabricius-McCoy comparison were verified. By this, we intend to make a detailed investigation about common features shown in those known models, compare the symmetry structure, then further explore more unknown models in the scheme of functional relations. Among the function relations, the fusion relation with "truncation" property plays a vital role in connecting the $T$- and Baxter's $Q$-matrix.  Usually one can derive the boundary fusion relation, i.e. the "truncation" identity, using the Baxter's TQ-relation for any $Q$-operator (not necessarily satisfying the $QQ$-relation) if a $Q$-matrix exists (see, e.g. \cite{Bax, Kor, PS}).  
Conversely, the validity of boundary fusion relation may as well  strongly suggests the existence of Baxter's $Q$-operator in many cases (see, e.g. \cite{Nep} and references therein).  The models we shall discuss in this article are the generic generalized $\tau^{(2)}$-model (where no $Q$-operator is known), superintegrable CPM, and XXZ spin chain of higher spin at the roots of unity. In the present paper, we obtain 
two main results related to these models. First, by constructing the fusion matrix from the explicit fused $L$-operator, we derive the fusion relations with the truncation identity for these models without relying on the theory of Baxter's $Q$-operator. Second, by the 
algebraic Bethe ansatz of XXZ spin chain of higher spin at the roots of unity and 
certain special $\tau^{(2)}$-models which include the superintegrable CPM, we show the  $sl_2$-loop-algebra symmetry of the root-of-unity XXZ chain for a higher spin as in the spin-$\frac{1}{2}$ case \cite{DFM, De05, FM01}. Furthermore, they all share the same structure as the superintegrable CPM  about the Bethe equation and the evaluation (Drinfeld) polynomial for the symmetry algebra as in \cite{R05b, R06Q}. As a consequence, a conjecture raised in \cite{NiD} about the simple polynomial property for the $sl_2$-loop-algebra evaluation parameters in the  XXZ chain of spin-$\frac{N-1}{2}$ has been justified. Note that for the $\tau^{(2)}$-model in the generic case where algebraic Bethe ansatz cannot be applied due to the lack of peudovacuum state, our fusion-matrix study strongly suggests, with computational evidences in cases, that the fusion relations with the truncation identity always hold. Hence by \cite{GIPS}, the separation-of-variables method provides the solution of Baxter equation associated to the $\tau^{(2)}$-model.

The quantum inverse scattering  method/algebraic Bethe ansatz developed by the Leningrad school in the early eighties  \cite{Fad, KBI, KS}  systematized earlier results about the Bethe ansatz of two-dimensional lattice models in an algebraic scheme by using the Yang-Baxter (YB) equation as a central role of solvability. 
A YB solution defines a local $L$-operator, which gives rise to the algebra of quantum monodromy matrices, called the ABCD-algebra. From the ABCD-algebra, one can construct a set of commuting transfer matrices, which in principal could be simultaneously diagonalized using a basis derived from the pseudovacuum state by the Bethe-ansatz technique. Furthermore, one can define the quantum determinant of the algebra, a concept first introduced in \cite{IK} (or see \cite{KBI} Chapter VIII), and played an important role in deriving the fused transfer matrices in this work. It is known that   
this algebraic method has long been used in the investigation of XXZ spin chains (see e.g., \cite{KiR, TakF} references therein), and in the $\tau^{(2)}$-model \cite{Ta, Ta92}. For the XXZ spin chain of higher spin in the root-of-unity case, it possesses some extra symmetry carrying the ''evaluation'' parameters, which indeed determine the eigenvalues of fusion matrices and implicitly encode symmetry of the model. Hence a new structure, not seen in the general XXZ chain of higher spin, appears in the root-of-unity theory. 
In this article, we employ the ABCD-algebra method in the generalized $\tau^{(2)}$-model and the root-of-unity XXZ chain of higher spin to study the fusion matrix through some explicit fused $L$-operators. The boundary fusion relation will be our main concern.  
The technique is first to make use of the quantum determinant of $L$-operators, not only on the  explicit form, but also its nature in commuting the fusion-product of elements so that the recursive fusion relation holds. Next
 the detailed analysis about "averaging" the $L$-operator leads to the boundary fusion relation.  Since the ABCD-algebra of generalized $\tau^{(2)}$-model carries a non-equivalent, though similar, structure as the algebra for the XXZ spin chain due to the non-symmetric Boltzmann weights in the $R$-matrix of $\tau^{(2)}$-model, 
we shall provide a more elaborate discussion about the algebraic Bethe ansatz of the $\tau^{(2)}$-model (though many like routine exercises in the field). This is because the correct formulation with the precise expression of physical quantities is non-trivial, and required for the later CPM algebraic-Bethe-ansatz discussion in this work when comparing it with the complete results of superintegrable CPM derived from the functional relations \cite{AMP, B90, B93, BBP}. It is 
also needed for the parallel symmetry discussion between superintegrable $N$-state CPM and the XXZ chain of spin-$\frac{N-1}{2}$. The algebraic Bethe ansatz is known to be applied to the superintegrable $\tau^{(2)}$-model;  however to what extent the results obtained by the algebraic-Bethe-ansatz method compared with the complete $\tau^{(2)}$-eigenvalues and its degeneracy known in the study of CPM  \cite{AMP, B93, R05o} has not been fully discussed in the literature to the best of the author's knowledge, especially about possible symmetry structures of the model. To this end, 
we propose a scheme for certain special classes of generalized $\tau^{(2)}$-model, with the superintegrable CPM included,  where the pseudovacuum state exists so that the  algebraic-Bethe-ansatz technique can be performed in the way like the root-of-unity XXZ spin chain.    
When applying to the superintegrable $\tau^{(2)}$-model, the setting enables us to conduct exact investigations of various problems. One can rediscover the Bethe equation, fusion relations, forms of eigenvalue spectrum and evaluation polynomials, known in the theory of CPM. Furthermore, certain eigenvectors derived from the pseudovacuum state can also be extracted by the algebraic Bethe ansatz method. Nevertheless, {\it only certain sectors} of the spectrum are covered by this scheme. In the case of the root-of-unity XXZ spin chain of higher spin, the algebraic Bethe ansatz method produces the correct form of evaluation polynomial for the degeneracy by a detailed analysis of the eigenvalues of fusion matrices. Then
the zero-averages of off-diagonal elements in the quantum monodromy matrix, corresponding to the vanishing property of the $N$-string creation operator, give rise to the $sl_2$-loop-algebra symmetry of the root-of-unity six vertex model by a "$q$-scaling" procedure in \cite{DFM}. Thereupon one can identify the evaluation polynomial of $sl_2$-loop-algebra representation for a Bethe state through the explicit Fabricius-McCoy current (\cite{FM04} (1.37)) of the model.

This paper is organized as follows. In section \ref{BBS}, we discuss the fusion relations of 
the generalized $\tau^{(2)}$-model. 
We begin  with some preparatory work in subsection \ref{ssec.alBBS} on the algebraic structure derived from YB relation for the generalized $\tau^{(2)}$-model \cite{Ta, Ta92}. Using standard techniques in the ABCD algebra and quantum determinant for the twisted $R$-matrix,  we construct in subsection \ref{ssec.FBBS} the fusion operators from the fused $L$-operators so that the recursive fusion relation holds. By studying the average of $L$-operators, we then show evidences, verified in cases by direct computations, that the boundary fusion relation is valid for the generalized $\tau^{(2)}$-model in subsection \ref{ss.bFuBBS}. 
In section \ref{sec:BABBS}, we study two special classes of BBS models, which include the superintegrable CPM,  by the algebraic-Bethe-ansatz method where the pseudo-vacuum exists. We then perform the investigation on the Bethe equation and Bethe states for such models in subsection \ref{ss.alBBS}. The algebraic-Bethe-ansatz discussion of special BBS models when restricted on the superintegrable $\tau^{(2)}$-model  recovers the Bethe equation and evaluation polynomial of Onsager-algebra symmetry in the superintegrable CPM \cite{AMP, B93, R05o}. The comparison of those algebraic-Bethe-ansatz results with the complete results known in the theory of superintegrable CPM is given in  subsection \ref{ss.alCPM}.
In section \ref{sec:6v}, we study the root-of-unity symmetry of XXZ spin chain with a higher spin. First we briefly review some basic concepts in the algebraic Bethe ansatz of XXZ spin chain that are needed for later discussions, (for more detailed information, see e.g.,  \cite{Fad, KS} references therein). Then we summarize results in \cite{Ka, Krp, KRS, Nep, R06Q, Sk} about the fusion relation for the spin-$\frac{1}{2}$ XXZ chain at roots of unity. Using the fused $L$-operators, we extend the construction of fusion operators in the spin-$\frac{1}{2}$ to the spin-$\frac{d-1}{2}$ XXZ chain at $N$th root of unity for $2 \leq d \leq N$ in subsection \ref{ssec.6FuBe}, where the fusion relations are derived. Furthermore through the  fusion-matrix-eigenvalue discussion, we extract the correct form of evaluation polynomial incorporated with the Bethe equation. In subsection \ref{ssec.6symd}, we show that the root-of-unity XXZ chain of spin-$\frac{d-1}{2}$ possesses the $sl_2$-loop-algebra symmetry, and verify the evaluation polynomial by the explicit Fabricius-McCoy current of Bethe states. In subsection \ref{ssec.6CPM}, we make the comparison between the root-of-unity XXZ chain of spin-$\frac{N-1}{2}$ and the $N$-state superintegrable CPM, which are known to be closely related in literature \cite{AMP, B93, NiD}. 
We close in section \ref{sec. F} with some concluding remarks.

\section{Fusion Relations and Algebraic Bethe Ansatz of Generalized $\tau^{(2)}$-model  \label{BBS}}
\setcounter{equation}{0}
We first briefly review some basic structures in the ABCD-algebra for the generalized $\tau^{(2)}$-model in subsection \ref{ssec.alBBS}.  Then in subsection \ref{ssec.FBBS}  we construct the fusion operators as the trace of fused $L$-operators so that the recursive fusion relation holds, and the boundary fusion relation will be discussed in subsection \ref{ss.bFuBBS}.

\subsection{ABCD-algebra and quantum determinant in the generalized $\tau^{(2)}$-model \label{ssec.alBBS}}
We start with some basic notions about algebraic structures in the generalized $\tau^{(2)}$-model. The summary will be sketchy, but also serve to establish notations, (for  more detailed information, see \cite{Ta} and references therein).

For a positive integer $N$, we fix the $N$th root of unity, $\omega= e^{\frac{2 \pi \sqrt{-1}}{N}}$. Denote by $\CZ^N $ the vector space of $N$-cyclic vectors with $\{ | n \rangle \}_{ n \in \ZZ_N}$ as the standard basis where $\ZZ_N = \ZZ/N\ZZ $ , and  $X, Z$ the  
$\CZ^N $-operators defined by $X |n \rangle = | n +1 \rangle$, $ Z |n \rangle = \omega^n |n \rangle $ for $n \in \ZZ_N$, which satisfy the Weyl relation, $XZ= \omega^{-1}ZX$, with $X^N=Z^N=1$. 
The $L$-operator of the generalized $\tau^{(2)}$-model 
is built upon the Weyl operators $X, Z$ with $\CZ^2$-auxiliary space and  $\CZ^N$-quantum space\footnote{Here we use the form of $L$-operator in accord with the convention used in \cite{BBP, R05o}, which is essentially the transpose of the $L$-operator in \cite{GIPS, Ta}.} :
\be
{\tt L} ( t ) =  \left( \begin{array}{cc}
        1 + t \kappa X  & ( \gamma - \delta X)Z \\
        t ( \alpha - \beta X)Z^{-1} & t \alpha \gamma + \frac{\beta \delta}{\kappa} X 
\end{array} \right) =: \left( \begin{array}{cc}
        A(t)  & B(t) \\
        C(t) &  D(t) 
\end{array} \right), \ \ t \in \CZ ,
\ele(G) 
where $\alpha, \beta, \gamma, \delta, \kappa \in \CZ$ are parameters, which satisfy the YB equation\footnote{Note that  (\req(G)) satisfy the YB relation (\req(YB)) as well for a general $\omega$ not necessary a root of unity, using the Weyl operators $X, Z$ with  $XZ= \omega^{-1}ZX$.}
\be
R(t/t') ({\tt L} (t) \bigotimes_{aux}1) ( 1
\bigotimes_{aux} {\tt L} (t')) = (1
\bigotimes_{aux} {\tt L}(t'))( {\tt L}(t)
\bigotimes_{aux} 1) R(t/t'),
\ele(YB)
for the $R$-matrix of the {\it asymmetric} six-vertex model,
\be
R(t) = \left( \begin{array}{cccc}
        t \omega - 1  & 0 & 0 & 0 \\
        0 &t-1 & \omega  - 1 &  0 \\ 
        0 & t(\omega  - 1) &( t-1)\omega & 0 \\
     0 & 0 &0 & t \omega - 1    
\end{array} \right).
\ele(Rtau)
Then the monodromy matrix for the quantum chain of size $L$,
\be
\bigotimes_{\ell=1}^L  {\tt L}_\ell (t) = {\tt L}_1(t) \otimes \cdots \otimes {\tt L}_L (t) =  \left( \begin{array}{cc} A_L(t)  & B_L (t) \\
      C_L (t) & D_L(t)
\end{array} \right), \ \ {\tt L}_\ell (t):= {\tt L}(t) \ {\rm at \ site} \ \ell,
\ele(monM)
again satisfy the YB equation (\req(YB)), and the $\omega$-twisted trace, 
$$\tau^{(2)}(t) := A_L(\omega t) + D_L(\omega t),$$
form a family of commuting operators of the $L$-tensor space $\stackrel{L}{\otimes} \CZ^N$ of $\CZ^N$.  We shall denote the spin-shift operator  of $\stackrel{L}{\otimes} \CZ^N$ again by $X (:= \prod_{\ell =1}^L X_\ell)$  if no confusion could arise, which carries the $\ZZ_N$-charge, denoted by $Q= 0, \ldots, N-1$. By 
\be
{[X , A_L ]} =  [X , D_L ] = 0 , \ \ X B_L = \omega^{-1} B_L X , \ \ X C_L = \omega C_L X ,
\ele(XAB)
$X$ commutes with the $\tau^{(2)}$-matrix.
 The relation (\req(YB)) for the monodromy matrix gives rise to an algebra structure of 
the operator-entries $A_L(t), B_L(t), C_L(t), D_L(t)$, called the ABCD-algebra, (algebra of quantum monodromy matrix), in which the following conditions hold:
\bea(lr)
[A(t), A(t')]= [B(t), B(t')]=[C(t), C(t')]=[D(t), D(t')]=0 ;& \\ 
( t \omega - t') A(t)B(t') = (t-t') B(t')A(t) + t (\omega-1) A(t')B(t) , & A, B \longrightarrow C, D ; \\
( t \omega - t') B(t)A(t') = (t-t')\omega A(t')B(t) + t' (\omega-1) B(t')A(t), & A, B \longrightarrow C, D ; \\
( t \omega - t') C(t')A(t) = (t-t') A(t)C(t') + t' (\omega-1) C(t)A(t'), & A, C  \longrightarrow B, D ; \\
( t \omega - t') A(t')C(t) = (t-t')\omega C(t)A(t') + t (\omega-1) A(t)C(t'), & A, C  \longrightarrow B, D .
\elea(ABCD)
The quantum determinant follows from the ABCD algebra by setting $t'= \omega t$:
\bea(rll)
& B(\omega t) A (t) = A(\omega t) B (t), & D(\omega t) C (t) = C (\omega t) D (t); \\ 
&A( t) C (\omega t) = \omega C(t) A (\omega t),& B(t) D(\omega t) = \omega D(t) B(\omega t), \\
{\rm det}_q (\bigotimes {\tt L}_\ell) (t) :=& D(\omega t) A(t) - C(\omega t) B(t) = & A(\omega t) D(t) - B(\omega t) C(t)  \\ 
=&  A(t)D(\omega t) - \omega C(t) B(\omega t) =& D(t)A(\omega t) - \omega^{-1}  B(t) C(\omega t) . 
\elea(YBc)
or equivalently, the quantum determinant of the monodromy matrix (\req(monM)) is characterized by rank-one property of 
$R( \omega^{-1})$ in the following relation,
\be
R(\omega^{-1}) (\otimes {\tt L}_\ell (t) \bigotimes_{aux}1) ( 1
\bigotimes_{aux} \otimes {\tt L}_\ell (\omega t)) = (1 \bigotimes_{aux} \otimes {\tt L}_\ell (\omega t))( \otimes {\tt L}_\ell (t)
\bigotimes_{aux} 1) R(\omega^{-1}) = {\rm det}_q (\otimes {\tt L}_\ell) (t) \cdot  R(\omega^{-1}),
\ele(qdet)
with the explicit form for ${\rm det}_q (\otimes {\tt L}_\ell) (t) $:
\be
{\rm det}_q (\otimes {\tt L}_\ell) (t)  = q (t)^L X^L , \  \ \ \ \ q (t):= \frac{\beta \delta}{\kappa} + (\alpha \delta + \omega \beta \gamma ) t + \omega \alpha \gamma \kappa t^2 .
\ele(detq)
The third- and fifth relations of (\req(ABCD)) yield
$$
\begin{array}{l}
A(t)B(s) = \frac{t- \omega s }{\omega (t-s)} B(s)A(t)+ \frac{(\omega-1)t}{\omega (t-s)}  B(t)A(s), \
D(t)B(s) = \frac{\omega  t-s }{\omega (t-s)} B(s)D(t) -  \frac{(\omega-1) t}{\omega (t-s)}  B(t) D(s) . 
\end{array}
$$
By moving $B(t_i)$'s to the left hand side of $A(t), D(t)$, one obtains 
\bea(ll)
A(t)\prod_{i=1}^m B(t_i) &= \prod_{i=1}^m \frac{ t- \omega t_i }{\omega (t-t_i)} \cdot \prod_{i=1}^m B(t_i) A(t) \\
& + \sum_{k=1}^m \frac{(\omega-1)t}{\omega (t-t_k )} \prod_{i=1, i \neq k }^m \frac{ t_k- \omega t_i }{\omega (t_k-t_i)} \cdot B(t)  \prod_{i=1, i \neq k }^m B(t_i) A(t_k), \\
D(t)\prod_{i=1}^m B(t_i) &= \prod_{i=1}^m \frac{\omega  t-t_i }{\omega (t-t_i)} \cdot \prod_{i=1}^m B(t_i)D(t) \\
&- \sum_{k=1}^m \frac{(\omega-1) t}{\omega (t-t_k)} \prod_{i=1, i\neq k}^m \frac{\omega t_k-t_i }{\omega (t_k-t_i)} \cdot B(t) \prod_{i=1, i \neq k }^m B(t_i) D(t_k) . 
\elea(ADbs)
Similarly the second- and fourth relations in (\req(ABCD)) yield 
\bea(ll)
A(t)\prod_{i=1}^m C(t_i) &= \prod_{i=1}^m \frac{\omega t  - t_i}{t-t_i } \cdot \prod_{i=1}^m C( t_i) A(t) \\
& - \sum_{k=1}^m  \frac{ (\omega-1)t_k }{t-t_k} \prod_{i=1, i \neq k }^m \frac{ \omega t_k  - t_i}{t_k -t_i} \cdot C(t)  \prod_{i=1, i \neq k }^m C(t_i) A(t_k), \\
D(t)\prod_{i=1}^m C(t_i) &= \prod_{i=1}^m \frac{ t- \omega t_i}{t-t_i} \cdot \prod_{i=1}^m C(t_i)D(t) \\
&+ \sum_{k=1}^m \frac{ (\omega-1)t_k}{t-t_k} \prod_{i=1, i\neq k}^m \frac{ t_k- \omega t_i}{t_k-t_i} \cdot C(t) \prod_{i=1, i \neq k }^m C(t_i) D(t_j) . 
\elea(ADcs)
Note that by scaling the $t$-variable, parameters in (\req(G)) can be reduced to the case $\alpha + \gamma =0$, among which, with one more constraint $\omega \beta + \delta =0$, one can express $-\alpha  = \gamma = y^{-1}$, $ - \beta =  \omega^{-1} \delta = \mu^2 x y^{-2}$, $- \kappa = \mu^2 y^{-2}$ for $(x, y, \mu ) \in \CZ^3$. For the $N$-state CPM,  the rapidity variables of $L$-operator (\req(G)) are defined by  
\be
k x^N  = 1 -  k'\mu^{-N}, \ \ \  k y^N  = 1 -  k'\mu^N, \ \ (x , y , \mu ) \in \CZ^3 ,
\ele(CPrap) 
where $k', k$ are temperature-like parameters with $k^2 + k'^2=1$. In the superintegrable case, the parameters in (\req(G)) and the quantum determinant are given by 
\be
-\alpha = - \beta = \gamma  =  \omega^{-1} \delta = -\kappa  = 1 ,  \ \ \ \ 
{\rm det}_q {\tt L} (t) = \omega h^2 (t) X , 
\ele(sCPM)
(see, e.g. \cite{R05o} Sect. 5).
Hereafter we shall always use $h(t)$  to denote 
\be 
h(t) : = 1 - t .
\ele(hfun)

\subsection{Fused $L$-operator in generalized $\tau^{(2)}$-model  \label{ssec.FBBS}}
Here we construct the fused $L$-operator ${\tt L}^{(j)}(t)$ for the fusion $\tau^{(j)}$-matrix with ${\tt L}^{(2)}(t)= {\tt L}(t)$ in (\req(G)). 

For convenience of notations, we shall also denote the standard basis $|\pm 1 \rangle$ of the $\CZ^2$-auxiliary space of ${\tt L}(t)$, and its dual basis  by $
\widehat{x}= |1 \rangle$ , $ \widehat{y}= |-1 \rangle $; $ x = \langle 1 |$ , $ y = \langle -1 |$. 
For non-negative integers $m, n $, we denote by $\widehat{x}^m \widehat{y}^n$ the completely symmetric $(m+n)$-tensor of $\CZ^2$ defined by
$$
{m+n \choose n} \widehat{x}^m \widehat{y}^n = \underbrace{\widehat{x}\otimes \ldots \otimes \widehat{x}}_{m} \otimes \underbrace{\widehat{y}\otimes \ldots \otimes \widehat{y}}_{n} + \ {\rm all \ other \ terms \ by \ permutations} , 
$$
similarly for $x^m y^n$. For $j \geq 1$, the $\CZ^j$-auxiliary space is the space of completely symmetric $(j-1)$-tensors of the $\CZ^2$, with the following canonical basis $e^{(j)}_k$ and the dual basis $e^{(j) *}_k$:
\be
e^{(j)}_k = \widehat{x}^{j-1-k} \widehat{y}^k, \ \ e^{(j) *}_k = {j-1-k \choose k} x^{j-1-k} y^k , \ \ k=0, \ldots, j-1.
\ele(Cjb)
The ${\tt L}^{(j)} (t)$ is the operator $\bigg( {\tt L}^{(j)}_{k, l} (t)\bigg)_{0 \leq k, l \leq j-1}$ with the $\CZ^j$-auxiliary and $\CZ^N$-quantum space, where ${\tt L}^{(j)}_{k, l} (t)$ is expressed by
\be
{\tt L}^{(j)}_{k, l} (t) = \langle e^{(j)*}_k | {\tt L}( \omega^{j-2}t) \otimes_{aux} \cdots \otimes {\tt L}(\omega t) \otimes_{aux} {\tt L}(t) | e^{(j)}_l \rangle .
\ele(Gj)
Then ${\tt L}^{(j)} (t)$ are intertwined by some $R^{(j)}$-matrix. With ${\tt L}^{(j)} (t)$ as the local operator, its monodromy matrix defines the commuting family of $\tau^{(j)}$-operators of $\stackrel{L}{\otimes} \CZ^N$,
\be
\tau^{(j)} (t) = {\rm tr}_{\CZ^j} (\bigotimes_{\ell=1}^L  {\tt L}^{(j)}_\ell ( \omega t)).
\ele(tauj)
We now show the fusion relation between $\tau^{(j+1)}$, $\tau^{(j)} $ and $\tau^{(j-1)} $ through the quantum determinant (\req(detq)).

Consider the auxiliary-space tensor $\CZ^2 \otimes \CZ^j$ as a subspace of $\stackrel{j+1}{\otimes} \CZ^2$ with the identification  
$$
e^{(j+1)}_{k+1} = \frac{1}{{j \choose k+1}} \bigg( {j-1 \choose k+1} \widehat{x} \otimes e^{(j)}_{k+1}  +
{j-1 \choose k} \widehat{y} \otimes e^{(j)}_k  \bigg) , \ \ k=-1, \ldots, j-1,  
$$ 
and denote $
f^{(j-1)}_k := \widehat{x} \otimes e^{(j)}_{k+1}  -  \widehat{y} \otimes e^{(j)}_k $ for $ 0 \leq k \leq j-2$.
Then $e^{(j+1)}_l, f^{(j-1)}_k$ form a basis of $\CZ^2 \otimes \CZ^j$ with the dual basis $e^{(j+1)*}_l, f^{(j-1)*}_k$ expressed by
$$
\begin{array}{ll}
e^{(j+1)*}_{k+1} = x \otimes e^{(j)*}_{k+1}  +  y \otimes e^{(j)*}_k , & f^{(j-1)*}_k = \frac{1}{{j \choose k+1}} \bigg( {j-1 \choose k} x \otimes e^{(j)*}_{k+1} - \ {j-1 \choose k+1}y \otimes e^{(j)*}_k   \bigg).
\end{array}
$$
Then the expression of $e^{(j+1)}_k, e^{(j+1)*}_l$ yields
\be
\langle e^{(j+1)*}_k | {\tt L}^{(j+1)} (t) | e^{(j+1)}_l \rangle = \langle e^{(j+1)*}_k | {\tt L} (\omega^{j-1}t) \otimes_{aux} {\tt L}^{(j)}(t) | e^{(j+1)}_l \rangle. 
\ele(Lj+1)
In order to determine the rest entries of ${\tt L} (\omega^{j-1}t) \otimes_{aux} {\tt L}^{(j)}(t)$, we need the following simple lemma.
\begin{lem}\label{lem:cd} 
The second equality in $(\req(qdet))$ is equivalent to the following relations:
\bea(ccl)
\langle x^2 | {\tt L}( \omega t) \otimes_{aux} {\tt L}(t) | \widehat{x} \wedge \widehat{y} \rangle &= \langle y^2 | {\tt L}( \omega t) \otimes_{aux} {\tt L}(t) | \widehat{x} \wedge \widehat{y} \rangle &= 0 , \\
\langle x \otimes y | {\tt L}( \omega t) \otimes_{aux} {\tt L}(t) | \widehat{x} \wedge \widehat{y} \rangle &= \langle y \otimes x | {\tt L}( \omega t) \otimes_{aux} {\tt L}(t) | - \widehat{x} \wedge \widehat{y} \rangle &= \frac{1}{2} {\rm det}_q {\tt L}(t) ,
\elea(qCom)
where $\widehat{x} \wedge \widehat{y} = \frac{1}{2} ( \widehat{x} \otimes \widehat{y} - \widehat{y} \otimes \widehat{x})$. 
Hence we have 
$$
\langle e^{(3)*}_k | {\tt L}( \omega t) \otimes_{aux} {\tt L}(t) | \widehat{x} \otimes \widehat{y} \rangle= \langle e^{(3)*}_k | {\tt L}( \omega t) \otimes_{aux} {\tt L}(t) | \widehat{y} \otimes \widehat{x} \rangle  \ \ {\rm for} \ k=0,1,2 .
$$
As a consequence, for an integer $j \geq 2$, and $v_i = \widehat{x}$ or $\widehat{y}$ for $1 \leq i \leq j-1$, we have 
\bea(ll)
&\langle e^{(j)*}_k | {\tt L}( \omega^{j-2}t) \otimes_{aux} \cdots \otimes_{aux} {\tt L}(\omega t) \otimes_{aux} {\tt L}(t) | v_1 \otimes v_2 \otimes \cdots \otimes v_{j-1} \rangle \\
= & \langle e^{(j)*}_k | {\tt L}( \omega^{j-2}t) \otimes_{aux} \cdots \otimes_{aux} {\tt L}(\omega t) \otimes_{aux} {\tt L}(t) | v_{\sigma_1} \otimes v_{\sigma_2} \otimes \cdots \otimes v_{\sigma_{j-1}} \rangle
\elea(gdet)
for $0 \leq k \leq j-1$, and all permutations $\sigma$. 
\end{lem} $\Box$ \par \vspace{.1in} \noindent
Since the entries of ${\tt L} ( \omega^{j-1}t) \otimes_{aux} {\tt L}^{(j)}( t)$ are determined by those of ${\tt L}( \omega^{j-1}t) \otimes_{aux} \cdots \otimes_{aux} {\tt L}(t)$, by (\req(gdet)) one has    
\be
\langle e^{(j+1)}_{k+1} | {\tt L} ( \omega^{j-1}t) \otimes_{aux} {\tt L}^{(j)}( t)  | f^{(j)}_{k'} \rangle 
=  0 , \ \ -1 \leq k \leq j-1, \ 0 \leq k' \leq j-2.
\ele(fzero)
Using (\req(qCom)) and (\req(gdet)), one finds
$$
\begin{array}{l}
\langle f^{(j-1)*}_k | {\tt L} ( \omega^{j-1}t) \otimes_{aux} {\tt L}^{(j)}( t)  | f^{(j-1)}_{k'} \rangle 
 = \langle f^{(j-1)*}_k | {\tt L} ( \omega^{j-1}t) \otimes_{aux} {\tt L}^{(j)}( t)  | (2 \widehat{x} \wedge \widehat{y}) \otimes \widehat{x}^{j-k'-2} \otimes \widehat{y}^{k'}   \rangle \\
= \frac{1}{{j \choose k+1}} \langle  ( {j-1 \choose k} x \otimes y - {j-1 \choose k+1}y \otimes x ) \otimes  e^{(j-1)*}_k     | {\tt L} (\omega^{j-1}t) \otimes_{aux} {\tt L}^{(j)}(t)  | (2 \widehat{x} \wedge \widehat{y}) \otimes \widehat{x}^{j-k'-2} \otimes \widehat{y}^{k'}     \rangle \\
=  {\rm det}_q {\tt L}(\omega^{j-2}t) \cdot \langle     e^{(j-1)*}_k  |  {\tt L}^{(j-1)}(t)  |  \widehat{x}^{j-k'-2} \otimes \widehat{y}^{k'}     \rangle ,
\end{array}
$$
which, by (\req(detq)) and (\req(qCom)), in turn yields
\be
\langle f^{(j-1)*}_k | {\tt L} ( \omega^{j-1}t) \otimes_{aux} {\tt L}^{(j)}( t)  | f^{(j-1)}_{k'} \rangle =   \langle e^{(j-1)*}_k  |  {\tt L}^{(j-1)}(t)  |  e^{(j-1)}_{k'} \rangle q(\omega^{j-2}t)X  .
\ele(fj-1)
From the definition (\req(tauj)) of $\tau^{(j)}$-matrices, the relations, (\req(Lj+1)) (\req(fj-1)) and (\req(fzero)), imply the following result:
\begin{prop}\label{prop:fusion} 
The $\tau^{(j)}$-matrices satisfy the (recursive) fusion relation by setting $\tau^{(0)}=0, \tau^{(1)} = I$,
\bea(l)
\tau^{(2)}(\omega^{j-1} t) \tau^{(j)}(t) =  z( \omega^{j-1} t) X \tau^{(j-1)}(t)  + \tau^{(j+1)}(t) , \ \ j \geq 1 .
\elea(fus)
where $z(t) = q(t)^L$ with $q(t)$ in $(\req(detq))$.
\end{prop}
$\Box$

\subsection{Boundary fusion relation in generalized $\tau^{(2)}$-model \label{ss.bFuBBS}}

For convenience, we introduce the following convention for a family of commuting operators $O(t)$:
$$
[O]_n (t) := \prod_{i=0}^{n-1} O(\omega^i t), \ \ n \in \ZZ_{\geq 0} , 
$$
and the average of $O(t)$ is defined by
$$
\langle O \rangle \ (= \langle O \rangle (t^N) ) = [O]_N (t). 
$$
The "classical" $L$-operator of BBS model is the average of (\req(G)):
\be
{\cal L} (t^N) = \left( \begin{array}{cc}
        \langle A \rangle   & \langle B \rangle \\
        \langle C \rangle &  \langle D \rangle 
\end{array} \right) = \left( \begin{array}{cc}
        1 + (-1)^{N+1}  \kappa^N t^N  & \gamma^N - \delta^N  \\
        (-1)^{N+1} ( \alpha^N - \beta^N )t^N &  \frac{\beta^N \delta^N}{\kappa^N} + (-1)^{N+1} \alpha^N \gamma^N t^N 
\end{array} \right) .
\ele(cG)
The averages $\langle A_L \rangle, \langle B_L \rangle, \langle C_L \rangle, \langle D_L\rangle$ of the monodromy matrix $(\req(monM))$ coincide with the classical $L$th monodromy associated to $(\req(cG))$ \cite{Ta}\footnote{The formula (\req(avM)) about averages of the monodromy matrix is stated in \cite{Ta} page 966 as a consequence of Proposition 5 (ii) there,  (or Lemma 1.5 in \cite{Ta92}), but without proof, and also not with the required form which could be misprint in both papers. The correct version should be $< \bigtriangleup(T) > = <T_1> <T_2>$, instead of $\bigtriangleup(<T>) = <T_1> <T_2>$. As the author could not find a proof in literature about the correct Tarasov's statement, here in this paper we provide a mathematical justification about the correct statement in Proposition \ref{prop:L(N+1)} and \ref{prop:6Va} for the generalized $\tau^{(2)}$-model and the XXZ spin chain respectively. }:
\be 
\left( \begin{array}{cc} \langle A_L \rangle  & \langle B_L \rangle \\
      \langle C_L \rangle & \langle D_L\rangle
\end{array} \right) (t^N) = {\cal L}_1 (t^N) {\cal L}_2 (t^N) \cdots {\cal L}_L (t^N) ( = {\cal L} (t^N)^L), 
\ele(avM)

By (\req(gdet)), the $(k, l)$th entry ${\tt L}^{(j)}_{k, l} (t)$ of ${\tt L}^{(j)}(t)$ is equal to the expression in (\req(gdet)) with $v_i = \widehat{x}$ for $1 \leq i \leq j-1-l$, and $\widehat{y}$ otherwise. Hence one can express ${\tt L}^{(j)}(t)$ in terms of entries in (\req(G)).
For example, the matrix-form of ${\tt L}^{(3)} (={\tt L}^{(3)}(t))$ is 
\be
 \left( \begin{array}{ccc}
         [A]_2(t), & A(\omega t) B(t),  & [B]_2(t) \\
        A(\omega t) C(t) + C(\omega t) A(t), & A(\omega t) D(t)+ C(\omega t) B(t),    & D(\omega t) B(t) + B(\omega t) D(t) \\
{[C]_2(t)}, & C(\omega t) D(t),  & [D]_2 (t)
\end{array} \right).
\ele(L3)
Note that one can also write ${\tt L}^{(3)}_{0,1} = A(\omega t) B(t)$,${\tt L}^{(3)}_{1,1} = D(\omega t) A(t)+ B(\omega t) C(t)$, ${\tt L}^{(3)}_{1,0} = D(\omega t) C(t)$. Among the ${\tt L}^{(j)}$-entries  for a general $j$, the following ones can be derived by setting  $\otimes v_i=\widehat{x}^{j-1-l} \otimes \widehat{y}^l$ in (\req(gdet)): 
\bea(ll)
{\tt L}^{(j)}_{0, 0} = [A]_{j-1}(t), ~ ~ {\tt L}^{(j)}_{0, j-1} = [B]_{j-1}(t), ~ ~ {\tt L}^{(j)}_{j-1, 0} = [C]_{j-1}(t), & {\tt L}^{(j)}_{j-1, j-1} = [D]_{j-1}(t); \\ 
{\tt L}^{(j)}_{0, l} = [A]_{j-l-1}(\omega^l t) [B]_l(t), ~ ~ {\tt L}^{(j)}_{j-1, l} = [C]_{j-l-1}(\omega^l t) [D]_l(t) , & 1 \leq l \leq j-2.   
\elea(Lj**)
By the relation, $e^{(j+1) *}_k = e^{(j) *}_{k-1} \otimes y + e^{(j) *}_k \otimes x$, for the dual basis of $\CZ^j$'s, one can compute ${\tt L}^{(j+1)}_{k,l}$ in terms of ${\tt L}^{(j)}$-entries  using the following recursive relations: 
\bea(l)
{\tt L}^{(j+1)}_{k, 0}(t) ={\tt L}^{(j)}_{k-1, 0}(\omega t) C(t) + {\tt L}^{(j)}_{k, 0}(\omega  t) A(t) , \\
{\tt L}^{(j+1)}_{k, l}(t) = {\tt L}^{(j)}_{k-1, l-1}(\omega t) D(t) + {\tt L}^{(j)}_{k, l-1}(\omega t) B(t) \ \ \ {\rm for} \ l \geq 1 , 
\elea(LRec)
by which, one can in principal derive the expression of ${\tt L}^{(j)}_{k, l}(t)$ from those in (\req(Lj**)) for a given $j$. However it is still not easy to obtain a close form of ${\tt L}^{(j)}_{k, l}$ for all $j$ except the following $(k, l)$'s:
\bea(ll)
{\tt L}^{(j)}_{1, 0}(t) = (\sum_{i=0}^{j-2} \omega^{j-i-2}  [A]_i (\omega^{j-i-1} t) [A]_{j-i-2} (\omega^{-1}t) )C (t),  & {\tt L}^{(j)}_{1, 0}, A, C \leftrightarrow {\tt L}^{(j)}_{j-2, j-1}, D, B ;  \\
{\tt L}^{(j)}_{j-2, 0}(t)=(\sum_{i=0}^{j-2} \omega^i A( \omega^{j-2i-2} t) ) [C]_{j-2} (t), & {\tt L}^{(j)}_{j-2, 0}, A, C \leftrightarrow {\tt L}^{(j)}_{1, j-1}, D, B . 
\elea(Ljfm)
Indeed, the above formulas are derived by the induction-method using the relations, $[C]_k(\omega t) A(t) = \omega^k A( \omega^{-k} t) [C]_k(t)$, $[B]_k(\omega t) D(t) = \omega^k D( \omega^{-k} t) [B]_k(t)$ for $k \geq 0$, which follows from the equalities
\be
C(t) A(t) = A( \omega^{-1} t) C(t), \ \ B(t) D(t) = \omega D( \omega^{-1} t) B(t).
\ele(CABD)  

In the study of CPM \cite{BBP}, the $\tau^{(j)}$-fusion relations (\req(fus)) with rapidity parameters in (\req(CPrap)) are truncated at $\tau^{(N+1)}$ with the following boundary fusion relation: 
\begin{eqnarray}
& \tau^{(N+1)}(t) = z(t ) X \tau^{(N-1)}( \omega t) + u (t) I,  & u(t):= (\langle A_L \rangle + \langle D_L \rangle )(t^N) , \label{fusB} 
\end{eqnarray}
where $A_L , D_L$ are operators in (\req(monM)). We are going to indicate the relation (\ref{fusB}) always holds in the generalized $\tau^{(2)}$-model for arbitrary parameters. 
Express the diagonal entries of $(L+1)$th monodromy matrix in (\req(monM))  by $
A_{L+1} (t) = (A_L \otimes A + B_L \otimes C)(t)$, $B_{L+1} (t) = (A_L \otimes B + B_L \otimes D)(t)$ , $(A_{L+1}, B_{L+1}, A_L,  B_L \rightarrow C_{L+1}, D_{L+1}, C_L,  D_L)$;
hence their averages: $ \langle A_{L+1} \rangle = [A_L \otimes A + B_L \otimes C]_N (t)$, $
\langle B_{L+1} \rangle = [ A_L \otimes B + B_L \otimes D)]_N(t)$, etc. 
Since $A_L, B_L$ and $C_L, D_L$ satisfy (\req(YBc)), using the expression (\req(Gj)) for  ${\tt L}^{(N+1)}_{k, 0}, {\tt L}^{(N+1)}_{k, N}$, one obtains  
\bea(ll)
\langle A_{L+1} \rangle  = \langle A_L \rangle \otimes \langle A \rangle + \langle B_L \rangle \otimes \langle C \rangle + \sum_{k=1}^{N-1} [A_L]_{N-k}(\omega^k t) [B_L]_k(t)\otimes {\tt L}^{(N+1)}_{k, 0}(t) , \\
\langle B_{L+1} \rangle  = \langle A_L \rangle \otimes \langle B \rangle + \langle B_L \rangle \otimes \langle D \rangle + \sum_{k=1}^{N-1} [A_L]_{N-k}(\omega^k t) [B_L]_k(t)\otimes {\tt L}^{(N+1)}_{k, N}(t) , \\
 A_{L+1}, B_{L+1}, A_L,  B_L \longrightarrow C_{L+1}, D_{L+1}, C_L,  D_L .
\elea(AL1)
Using the above relations and the following proposition on the vanishing ${\tt L}^{(N+1)}_{k, 0}, {\tt L}^{(N+1)}_{k, N}$ for $k \neq 0, N$, one can determine the average $\langle A_L \rangle$, $\langle B_L \rangle$ etc.  through the relation (\req(avM)).
\begin{prop}\label{prop:L(N+1)} The entries ${\tt L}^{(N+1)}_{k, 0}, {\tt L}^{(N+1)}_{k, N}$ of ${\tt L}^{(N+1)}(t)$ (on the 1st and $N$th column) are all zeros except $
{\tt L}^{(N+1)}_{0, 0} = \langle A \rangle$, ${\tt L}^{(N+1)}_{N, 0} = \langle C \rangle$, ${\tt L}^{(N+1)}_{0, N} = \langle B \rangle$, ${\tt L}^{(N+1)}_{N, N} = \langle D \rangle$. As a consequence, the relation $(\req(avM))$ holds; hence  $u(t)$ in $(\ref{fusB})$ is equal to $ {\rm tr} ( {\cal L} (t^N)^L ) $ with ${\cal L} (t^N)$ in $(\req(cG))$.
\end{prop}
{\it Proof.} We need only to show the vanishing of ${\tt L}^{(N+1)}_{k, 0}, {\tt L}^{(N+1)}_{k, N}$ for $k \neq 0, N$. 
The relation (\req(AL1)) with $L=1$  yields  
$$
\begin{array}{l}
\langle A_2 \rangle  = \langle A \rangle^2 + \langle B \rangle \langle C \rangle + \sum_{k=1}^{N-1} [A]_{N-k}(\omega^k t) [B]_k(t)\otimes {\tt L}^{(N+1)}_{k, 0}(t) , \\
\langle B_2 \rangle  = \langle A \rangle \langle B \rangle + \langle B \rangle  \langle D \rangle + \sum_{k=1}^{N-1} [A]_{N-k}(\omega^k t) [B]_k(t)\otimes {\tt L}^{(N+1)}_{k, N}(t), 
\end{array}
$$
which are operators invariant when changing $t$ by $\omega t$.
Note that $ X^i Z^k$ $( i, k \in \ZZ_N)$ form a $\CZ$-basis of all $N$-by-$N$ matrices; and the term  $[A]_{N-k}(\omega^i t) [B]_k(\omega^j t)$ in the above expressions can be written in the form  $\sum_{m=0}^{N-1} p_m(t) X^m Z^k$ for some $p_m(t) \in \CZ[t]$. By this, each term appeared in the above expression of $\langle A_2 \rangle, \langle B_2 \rangle$ is again invariant under $t \mapsto \omega t$, i.e.,  for $1 \leq k \leq N-1$ and $e=0, N$,  the equality holds:
$$
[A]_{N-k}(\omega^{k+1} t) [B]_k(\omega t)\otimes {\tt L}^{(N+1)}_{k, e}(\omega t) = [A]_{N-k}(\omega^k t) [B]_k(t)\otimes {\tt L}^{(N+1)}_{k, e}(t).
$$
This implies $
A(t) \otimes {\tt L}^{(N+1)}_{k, e}(\omega t) = A(\omega^k t) \otimes {\tt L}^{(N+1)}_{k, e}(t)$ 
by the relation, $
A(\omega^k t) [A]_{N-k}(\omega^{k+1} t) [B]_k(\omega t) =  A(t) [A]_{N-k}(\omega^k t) [B]_k(t)$. The linear independence of matrices $1$ and $X$ in the expression of $A(t)$ in turn yields 
$$
 {\tt L}^{(N+1)}_{k, e}(\omega t) =  {\tt L}^{(N+1)}_{k, e}(t) = \omega^k {\tt L}^{(N+1)}_{k, e}(t) . 
$$
Therefore ${\tt L}^{(N+1)}_{k, 0}(t) = {\tt L}^{(N+1)}_{k, N}(t)  =0$ for $1 \leq k \leq N-1$.
$\Box$ \par \noindent \vspace{.1in}
{\bf Remark}. For the superintegrable $\tau^{(2)}$-model (\req(sCPM)), by (\req(cG)), the above proposition implies the averages of $L$th monodromy matrix-entries are: $\langle A_L \rangle = \langle D_L \rangle = (1- t^N)^L$,  $\langle B_L \rangle = \langle C_L \rangle = 0$. In CPM where the rapidities $p, p'$  of $\tau^{(2)}$-model are in (\req(CPrap)) with $p$ fixed and $t= x_{p'}y_{p'}$ for $p'$, one can compute the eigenvalues of (\req(cG)) using the coordinate $\lambda = \mu^N$ to express the $u(t)$ in (\ref{fusB}) for 
CPM: $u(t) = \alpha_p (\lambda_{p'})+ \alpha_p(\lambda_{p'})$ where $\alpha_p (\lambda_{p'}) = ( \frac{k'(1-\lambda_p \lambda_{p'})^2 }{\lambda_{p'}(1- k' \lambda_p)^2})^L $ ( \cite{BBP} (4.27)).

By Proposition \ref{prop:L(N+1)}, the $(N+1)$th fused $L$-operator ${\tt L}^{(N+1)}$ can be written in the form
$$
{\tt L}^{(N+1)} (t) = \left( \begin{array}{ccc}
         \langle A \rangle  & * & \langle B \rangle \\
        0 & \widetilde{\tt L}^{(N-1)}(t) &  0 \\
     \langle C \rangle & *'  &\langle D \rangle   
\end{array} \right)
$$
where $* = ({\tt L}^{(N+1)}_{0,1}, \cdots, {\tt L}^{(N+1)}_{0,N-1})$, $*'= ({\tt L}^{(N+1)}_{N,1}, \cdots, {\tt L}^{(N+1)}_{N,N-1})$, $\widetilde{\tt L}^{(N-1)}(t) = \bigg( \widetilde{\tt L}^{(N-1)}_{k, l}(t) \bigg)_{0 \leq k, l \leq N-2}$ with $\widetilde{\tt L}^{(N-1)}_{k, l}(t) := {\tt L}^{(N+1)}_{k+1, l+1} (t)$. The boundary fusion relation $(\ref{fusB})$ is equivalent to the relation between $\tau^{(N-1)}$-matrix and the $L$th monodromy matrix for $\widetilde{\tt L}^{(N-1)}$:
\be
z(t ) X \tau^{(N-1)}( \omega t) = {\rm tr}_{\CZ^{N-1}} (\bigotimes_{\ell=1}^L  \widetilde{\tt L}^{(N-1)}_\ell ( \omega t)).
\ele(TiTau)
The above relation (\req(TiTau)) will hold if we can prove $\widetilde{\tt L}^{(N-1)}(\omega t)$ similar to ${\tt L}^{(N-1)}(\omega^2 t)\cdot {\rm det}_q {\tt L}(t)$ by a diagonal similarity transformation $M$ (independent to $t$): 
\be
\widetilde{\tt L}^{(N-1)}(\omega t) =  M {\tt L}^{(N-1)}(\omega^2 t) M^{-1} \cdot {\rm det}_q {\tt L}(t).
\ele(LGaug)
We now determine the above similarity relation for $N=2,3,4$. 
For $N=2$ where $\omega =-1$, by (\req(L3)) one has $\widetilde{\tt L}^{(1)}(\omega t)= A(t) D(\omega t)- \omega C(t) B(\omega t)= {\rm det}_q {\tt L}(t)$, hence follows (\req(TiTau)).
For $N=3$, we use (\req(gdet)) to express ${\tt L}^{(N+1)} $-entries, 
$$
\begin{array}{l}
\widetilde{\tt L}^{(2)}_{0,0}(\omega t) = A(t) A(\omega^2 t) D(\omega t) + A(t) C(\omega^2 t) B(\omega t)+ C(t) A(\omega^2 t) B(\omega t) , \\
\widetilde{\tt L}^{(2)}_{0,1}(\omega t) = B(t) B(\omega^2 t) C(\omega t) + B(t) D(\omega^2 t) A(\omega t)+ D(t) B(\omega^2 t) A(\omega t) , \\
\widetilde{\tt L}^{(2)}_{1,0}(\omega t) = A(t) C(\omega^2 t) D(\omega t) + C(t) A(\omega^2 t) D(\omega t)+ C(t) C(\omega^2 t) B(\omega t), \\
\widetilde{\tt L}^{(2)}_{1,1}(\omega t) = B(t) D(\omega^2 t) C(\omega t) + D(t) B(\omega^2 t) C(\omega t)+ D(t) D(\omega^2 t) A(\omega t) . 
\end{array}
$$
By (\req(CABD)) and (\req(detq)), one arrives  
$$
\begin{array}{ll}
\widetilde{\tt L}^{(2)}_{0,0}(\omega t) = A(\omega^2 t) A(t) D(\omega t) + (\omega A(t) +  \omega^2  A(\omega t) ) \omega C(t)  B(\omega t) &= A(\omega^2 t) {\rm det}_q {\tt L}(t), 
\\ \widetilde{\tt L}^{(2)}_{0,1}(\omega t) 
= B(\omega^2 t) B(t) C(\omega t) + B(t) (D(\omega^2 t) + \omega^2 D(\omega t))  A(\omega t) &= - \omega B(\omega^2 t) {\rm det}_q {\tt L}(t) , \\
\widetilde{\tt L}^{(2)}_{1,0}(\omega t) 
= C(\omega^2 t) ( A(\omega t)  + \omega A(\omega^2 t)) D(\omega t)+ C(\omega^2 t) C(t)  B(\omega t)  &= - \omega^{-1}C(\omega^2 t) {\rm det}_q {\tt L}(t), \\
\widetilde{\tt L}^{(2)}_{1,1}(\omega t) 
= (\omega D(\omega t)   + D(t) ) B(t) C(\omega t)+ D(\omega^2 t) D(t)  A(\omega t)
&= D(\omega^2 t){\rm det}_q {\tt L}(t) . 
\end{array}
$$
Hence (\req(LGaug)) holds with $M = {\rm dia}[ -\omega , 1]$. Indeed, the same method as in above arguments, together with (\req(Lj**)) and (\req(Ljfm)), yields the following four relations for an arbitrary $N$ with $\omega^N=1$ :
\bea(ll)
\widetilde{\tt L}^{(N-1)}_{0,0}(\omega t) = {\tt L}^{(N-1)}_{0,0}(\omega^2 t) {\rm det}_q {\tt L}(t) , &
\widetilde{\tt L}^{(N-1)}_{0,N-2}(\omega t)  = - \omega {\tt L}^{(N-1)}_{0,N-2}(\omega^2 t) {\rm det}_q {\tt L}(t) , \\
\widetilde{\tt L}^{(N-1)}_{N-2,0}(\omega t) = - \omega^{-1}{\tt L}^{(N-1)}_{N-2,0}(\omega^2 t) {\rm det}_q {\tt L}(t), &
\widetilde{\tt L}^{(N-1)}_{N-2,N-2}(\omega t) ={\tt L}^{(N-1)}_{N-2,N-2}(\omega^2  t){\rm det}_q {\tt L}(t) . 
\elea(L4)

For $N=4$ where $\omega = \sqrt{-1}$, using the explicit form of $A, B, C, D$ in (\req(G)) to compute the $\widetilde{\tt L}^{(3)}$-entries other than those in (\req(L4)), then comparing $\widetilde{\tt L}^{(3)}$ with (\req(L3)), we can verify the relation (\req(LGaug)) holds with $M = {\rm dia}[ -\sqrt{-1} , \frac{1-\sqrt{-1}}{2}, 1]$. For an arbitrary given $N$, the similar transformation in (\req(LGaug)) could be obtained by direct calculation, however the general structure has not been found yet.

\section{Algebraic Bethe Ansatz in Generalized $\tau^{(2)}$-model \label{sec:BABBS}}
\setcounter{equation}{0}
In this section, we use the algebraic-Bethe-ansatz techniques to discuss certain BBS models centered at the superintegrable $\tau^{(2)}$-model, where the pseudovacuum state exists.

First we note that the non-trivial kernel of the operator $C(t)$ in (\req(G)) is equivalent to $\langle C \rangle = 0$. By scaling the $t$-variable  and changing $X$ by $\omega^k X$, one may set $\alpha = \beta = -1$, i.e.,  
\be
{\tt L} ( t ) =  \left( \begin{array}{cc}
        A(t)  & B(t) \\
        C(t) &  D(t) 
\end{array} \right) = \left( \begin{array}{cc}
        1 + t \kappa X  & ( \gamma - \delta X)Z \\
       -t(1-X)Z^{-1} & - t  \gamma - \frac{ \delta}{\kappa} X 
\end{array} \right) . 
\ele(sBBS) 
In this section we shall consider only the case (\req(sBBS)) with the following constraint:
\be
\gamma \neq \omega^{-k} \delta \ \ \  \ \ {\rm for} \ k=2, \ldots, N, 
\ele(gadel)
which will be called the special BBS model in this paper.  
The quantum determinant in (\req(detq)) now takes the form 
\be
{\rm det}_q {\tt L}(t) = q (t) X , \ \ q (t) = \omega h_1(t) h_2(t), \ \ {\rm where} \ h_1 (t):= (1 +  \kappa t) , \  h_2 (t) := (- \gamma t - \frac{ \omega^{-1} \delta}{\kappa}) .
\ele(h12)
The special BBS model with one further constraint $\langle A \rangle = \langle D \rangle$, equivalently  $
\gamma^N = \delta^N = (-\kappa)^N  $, implies  $\gamma = \omega^{-1} \delta =  - \omega^{-i_0} \kappa$ for some $i_0$. By substituting $ \gamma t$ by $t$, then applying a similar transformation, ${\tt L}(t)$ is equivalent to the case  $\gamma =1$, with 
$B(t) = (1 - \omega X)Z$, $A(t) = 1 - \omega^{i_0} t X$, $ D(t) = - t   + \omega^{i_0} X$, and $  h_1 (t) = \omega^{i_0 } h_2 (t) =   h (t) $, where $h(t)$ is in (\req(hfun)). 
In particular, when $i_0=0$,  one arrives the superintegrable $\tau^{(2)}$-model (\req(sCPM)).

\subsection{Bethe equation of the special BBS Model  \label{ss.alBBS}}
Denote the eigenvectors of the $\CZ^N$-operator $X$ by  
$$
f_k := Z^k (\sum_{n \in \ZZ_N} |n \rangle), \ \ X (f_k) = \omega^{-k} f_k  \ \ {\rm for}  \ \ k \in \ZZ_N . 
$$
Let $A_L, B_L, C_L, D_L$ be the entries of the $L$th monodromy matrix (\req(monM)) with ${\tt L}(t)$ in (\req(sBBS)) satisfying the condition (\req(gadel)), and the transfer matrix is given by $\tau^{(2)}( \omega^{-1} t) = A_L(t)+ D_L (t)$. 
It is easy to see the  $C(t)$ in (\req(sBBS)) has one-dimensional kernel space generated by $v:= f_1$, and $
A(t)v = h_1 (\omega^{-1} t)v$, $D(t) = h_2 (t)v$, with $h_1, h_2$ in (\req(h12)). Define the pseudo-vacuum:
$$
\Omega_L = \otimes^L_{\ell =1} v_\ell \in \stackrel{L}{\otimes} \CZ^N , \ \ v_\ell = v \ {\rm  for \ all } \ \ell,
$$
then $A_L(t)  \Omega_L = h_1 ( \omega^{-1} t)^L \Omega_L$ , $
D_L(t)  \Omega_L = h_2 (t)^L \Omega_L$, $ C_L(t) \Omega_L = 0$ and  
\bea(l)
B_L(t) \Omega_L =(\gamma - \delta \omega^2) \sum_{\ell=1}^L h_1(\omega^{-1} t)^{\ell-1} h_2(t)^{L-\ell} (1 \otimes \cdots \otimes Z_{\ell th} \otimes \otimes \cdots \otimes 1) \Omega^L .
\elea(BVac)
Note that for $L=1$, we have $\Omega_L = f_1$ and $B_L(t) = B = (\gamma - \delta X)Z$. Then $B^k \Omega_L  = \omega^{-k} \prod_{i=2}^k (\gamma - \delta \omega^{-i-1})f_{k+1}$ for $0 \leq k \leq N-1$ are non-zero vectors by (\req(gadel)), which form a basis of $\CZ^N$.

For $m$ distinct non-zero complex numbers $t_j, 1 \leq j \leq m$, we consider the vector 
$$
\Psi (t_1, \ldots , t_m ) := \prod_{i=1}^m B_L (t_i) \Omega_L \in  \stackrel{L}{\otimes} \CZ^N ,
$$
and define the $t$-function,  
\be
\Lambda (t; t_1, \ldots, t_m)
= h_1(\omega^{-1} t )^L  \frac{F( \omega^{-1} t)}{F(t)}  +  h_2(t)^L \omega^{-m}  \frac{F(\omega  t )}{F(t)}, \ \ F(t) := \prod_{i=1}^m (1- t_i^{-1} t) .
\ele(LamBBS)
The polynomial-criterion of $\Lambda (t; t_1, \ldots, t_m)$ is given by the following  Bethe equation for $t_i$'s:
\be
\frac{h_2( t_i)^L }{h_1(\omega^{-1} t_i)^L }  = -  \omega^m \frac{F( \omega^{-1} t_i)}{F( \omega t_i) },
\ \ \ i = 1, \ldots, m .
\ele(BeBBS)
By (\req(XAB)), the $\ZZ_N$-charge of pseudo-vacuum and Bethe vectors are 
\be
X \Omega_L = \omega^{-L} \Omega_L , \ \ X \Psi (t_1, \ldots , t_m ) = \omega^{-L-m}\Psi (t_1, \ldots , t_m ) .
\ele(QBeV)
\begin{prop}\label{prop:BBSBe} 
The polynomial condition of $\Lambda (t; t_1, \ldots, t_m)$ is equivalent to 
the Bethe equation $(\req(BeBBS))$ for $t_i$'s, which is the necessary and sufficient condition for $\Psi (t_1, \ldots , t_m )$ as a common eigenvector of the transfer matrices $\tau^{(2)}( \omega^{-1} t)$. In this situation, $\Lambda (t; t_1, \ldots, t_m)$ is the eigen-polynomial for the eigenvector $\Psi (t_1, \ldots , t_m )$. 
\end{prop}
{\it Proof.} Write the function (\req(LamBBS)) in the form
$$
\Lambda (t; t_1, \ldots, t_m)= h_1 (\omega^{-1} t )^L \prod_{i=1}^m \frac{( t- \omega t_i )}{\omega (t-t_i)}  +  h_2(t )^L \prod_{i=1}^m \frac{\omega  t-t_i }{\omega (t-t_i)} .
$$
By (\req(ADbs)), $(A_L(t)+D_L(t)) \prod_{i=1}^m B_L (t_i) \Omega_L$  is equal to the sum of $\Lambda (t; t_1, \ldots, t_m) \Psi (t_1, \ldots , t_m )$ and $\Lambda_k (t; t_1, \ldots, t_m) B(t) \prod_{i=1, i \neq k }^m B_L (t_i) \Omega_L$, $(k=1, \ldots, m)$ where 
$$
\Lambda_k (t; t_1, \ldots, t_m) := \frac{(\omega-1)t}{\omega (t-t_k)}   \bigg( h_1( \omega^{-1} t_k )^L   \prod_{i \neq k} \frac{( t_k- \omega t_i )}{\omega (t_k-t_i)} - h_2
( t_k )^L \prod_{i \neq k}\frac{\omega  t_k- t_i }{\omega (t_k- t_i)} \bigg).
$$ 
By this,  $\Psi (t_1, \ldots , t_m )$ is a common eigenvector of $\tau^{(2)}( \omega^{-1} t)$ provided that $\Lambda_k (t; t_1, \ldots, t_m) =0$ for all $k$, which is equivalent to the Bethe equation (\req(BeBBS)). Then follows the proposition. 
$\Box$ \par \noindent \vspace{.1in}
{\bf Remark} In the above algebraic-Bethe-ansatz discussion of BBS model, we set $C(t) = -t(1-X)Z^{-1}$, which is the one appeared in superintegrable CPM. One may also set $B(t) = (1- \omega X)Z$ instead, and let $C(t)= t(\alpha - \beta X)Z^{-1}$ with parameters $\alpha, \beta$; then conduct a similar algebraic-Bethe-ansatz discussion. In the latter setting, the pseudo-vacuum $\Omega'_L = \otimes (f_0)_\ell$ satisfies the relations, $B_L(t) \Omega'_L  = 0$, $ A_L(t) \Omega'_L  = h_1(t)^L \Omega'_L $, and $ D_L(t)\Omega'_L  = \omega h_2(\omega^{-1} t)\Omega'_L$. With the same argument in Proposition \ref{prop:BBSBe} but using the relation (\req(ADcs)) instead, 
one considers the Bethe vector $\Psi'_L (t_1, \ldots , t_m ) := \prod_{i=1}^m C_L (t_i) \Omega'_L$, with the Bethe equation
\be
 \frac{h_1(t)^L }{h_2(\omega^{-1} t)^L} = - \omega^{L+m} \frac{F( \omega ^{-1} t)}{F( \omega t  )},  
\ele(BeBBS')
which in turn yields the eigenvalue expression for $\Psi'_L (t_1, \ldots , t_m )$: 
\be
\Lambda' (t; t_1, \ldots, t_m)
=  \omega^{L+m} h_2(\omega^{-1} t)^L \frac{F( \omega ^{-1} t)}{F( t )} + h_1(t)^L \frac{F( \omega t  )}{F(t)}.
\ele(Lam')
$\Box$ \par \vspace{.2in}

We now consider the special BBS model (\req(sBBS)) in $N=2$ case where $\omega=-1$, and the Bethe equation (\req(BeBBS)) becomes
$$
(\frac{\gamma t_i -   \delta/\kappa}{\kappa t_i -1})^L  = (- 1)^{m+1},   \ \ i = 1, \ldots, m .
$$
The solutions of the above relation are determined by non-zero roots of  $(\gamma t -   \delta/\kappa)^L  = \pm (\kappa t -1)^L$, each has $L$ solutions when $\gamma^L \neq  \pm \kappa^L \neq  \delta^L$:
$$
\begin{array}{lll}
r_k = \frac{ \delta  - \kappa \sigma^k_L}{\kappa(\gamma  - \sigma^k_L \kappa)} , &   r'_k =  \frac{\delta  - (-1)^{1/L} \sigma^k_L }{\kappa (\gamma  - (-1)^{1/L} \sigma^k_L  \kappa)}  &   (1 \leq k \leq L), \ \ \sigma_L := e^{\frac{2 \pi \sqrt{-1}}{L }} .
\end{array}
$$
For generic $\gamma , \kappa$ and $\delta$, the Bethe states are expected to generate the quantum space $\stackrel{L}{\otimes} \CZ^2$,  e.g., for $L=2$, it follows from the expression of the Bethe states $\Omega_2 (= f_1 \otimes f_1)$, $\Psi(r_1), \Psi(r_2)$, $\Psi(r'_1, r'_2)$:
$$
\begin{array}{cl}
\frac{1}{\gamma -  \delta } \Psi (r_i)  =&   h_1(- r_i) f_1 \otimes f_0 + h_2(r_i) f_0 \otimes f_1 ,  \ \ (i=1,2) ; \\
\frac{1}{(\gamma -  \delta)^2 } \Psi(r'_1, r'_2)=& \frac{ \gamma + \delta}{\gamma -  \delta}(h_1( - r'_1) h_1( -r'_2) + h_2(r'_1) h_2(r'_2))\Omega_2 +  ( h_1( r'_1) h_2(r'_2) - h_2( -r'_1 ) h_1( -r'_2) )  f_0 \otimes f_0 . 
\end{array}
$$
Note that in the superintegrable case (\req(sCPM)), $\Omega_2$ is the only Bethe state  in the above setting. Indeed in this situation, $\tau^{(2)}(-t) = (1 + t^2) (1 +  X \otimes X)-2t(Z \otimes Z) (1-X \otimes X )$, with the eigenvalues $2(1 + t^2), \pm 4t$, and eigenspaces $\langle \Omega_2, f_0 \otimes f_0 \rangle$, $ \langle  f_0 \otimes f_1 \mp  f_1 \otimes f_0 \rangle$ respectively.

For $N \geq 3$, it is not easy to obtain the complete solutions of the Bethe equation (\req(BeBBS)) in general, hence difficult to determine the space generated by Bethe states in $\stackrel{L}{\otimes} \CZ^N$. For $m=1$, the Bethe equation (\req(BeBBS)) becomes $
h_2( t )^L = h_1(\omega^{-1} t)^L $, 
which has $L$ solutions in generic cases. By (\req(BVac)), the Bethe states for $m=1$ likely form a basis for the subspace generated by $(1 \otimes \cdots \otimes Z_{\ell th} \otimes \otimes \cdots \otimes 1) \Omega^L $ for all $\ell$.

\subsection{Algebraic Bethe ansatz in superintegrable chiral Potts model \label{ss.alCPM}}
We now discuss the superintegrable $\tau^{(2)}$-model (\req(sCPM)) in the setting of  previous subsection, then compare the result with those in the theory of superintegrable CPM  \cite{AMP, B93}. 

In the study  of the chiral-Potts-transfer matrix in superintegrable CPM, there are quantum numbers, $P_a$ and $P_b$,  appeared in the eigenvalues, which satisfy the conditions
$$
0 \leq r ~ (:= P_a+P_b) \leq N-1, ~  \ P_b-P_a \equiv Q+L \pmod{N},   ~  \ LP_b \equiv m (Q-2P_b-m) \pmod{N} 
$$
with $Q$ the $\ZZ_N$-charge as before ((C.3) (C.4) in \cite{AMP}, (6.16) in \cite{B93}, or (59) (63) in \cite{R05o}).
The Bethe equation of superintegrable CPM is given by ((4.4) in \cite{AMP}, (6.22) in \cite{B93}):
\be
\frac{H_{\rm CP} ( \omega^{-1} t_i)^L }{H_{\rm CP} ( t_i)^L }  = -  \omega^{-r} \frac{F( \omega^{-1} t_i)}{F( \omega t_i) }
\ \ \ ( i = 1, \ldots , m ) ,  \ \ F(t) := \prod_{i=1}^m (1- t_i^{-1} t), 
\ele(BeCP)
where $H_{\rm CP} (t)$ is the polynomial defined by 
\be
H_{\rm CP} (t) = \frac{1- t^N}{1-t} ( = \frac{1- t^N}{h (t)}) .
\ele(HCP)
Define the normalized $\tau^{(2)}$- and $\tau^{(j)}$-matrices (\cite{R06Q} (3.4)) by setting $T^{(0)} = 0$, $T^{(1)}( t) = H_{\rm CP} ( \omega^{-1}t)^L$, and
\be
T^{(2)}(t)= \frac{\omega^{-P_b} (1- t^N)^L \tau^{(2)}(\omega^{-1} t) }{(1- \omega^{-1} t)^L(1-  t)^L}, \ \ T^{(j)}( t) = \frac{\omega^{-(j-1)P_b} (1- t^N)^L \tau^{(j)}(\omega^{-1}t) }{\prod_{k=-1}^{j-2}(1- \omega^k t)^L} .
\ele(CPTj)  
Then $T^{(j)}$-eigenpolynomials determined by $F(t)$ in (\req(BeCP)) satisfy the relations (\cite{R06Q} (3.8)-(3.11)):
\bea(l)
T^{(2)}(t)  F( t) = \omega^{-r} H_{\rm CP} (t)^L F( \omega^{-1}t)  +  H_{\rm CP} ( \omega^{-1} t)^L F( \omega t), \\    
T^{(j)}(t) = F( \omega^{-1} t) F( \omega^{j-1}t) \sum_{k=0}^{j-1} \frac{H_{\rm CP}(\omega^{k-1}t)^L \omega^{-kr} }{F(\omega^{k-1}t)F(\omega^k t)}, \ \ \ \ ( j \geq 1 ) , \\
T^{(j)}( t) T^{(2)}(\omega^{j-1} t) = \omega^{-r} H(\omega^{j-1} t)^L T^{(j-1)}(t)  + H(\omega^{j-2}t)^L T^{(j+1)}(t),  \ \ ( j \geq 1 ) , \\
T^{(N+1)}(t) = \omega^{-r} T^{(N-1)}( \omega t)
  + 2 H ( \omega^{-1}t)^L .
 \elea(TFfus)
Note that the above $T^{(j)}$-relations  are equivalent to the fusion relations (\req(fus)) (\ref{fusB}) for $\tau^{(j)}$'s in the superintegrable $\tau^{(2)}$-model ((53) in \cite{R05o}). The second relation of (\req(TFfus)) in turn yields
$$
T^{(N)}(t) = \omega^{P_b} F( \omega^{-1} t) F( \omega^{N-1}t) \sum_{k=0}^{N-1} {\cal P}_{\rm CP} (t^N)
$$
where ${\cal P}_{\rm CP} (t^N)$ is the evaluation polynomial of the Onsager-algebra representation for the degenerate eigenspaces in the superintegrable CPM \cite{R05o}:
\be
{\cal P}_{\rm CP} (t^N) = \omega^{-P_b} \sum_{k=0}^{N-1} \frac{H_{\rm CP}(t)^L (\omega^k t)^{-P_a-P_b}}{F(\omega^k t) F(\omega^{k+1} t)}.
\ele(pCP)
Bethe equation (\req(BeBBS)) in the superintegrable $\tau^{(2)}$-model  is the same as (\req(BeCP)), but with the constraint $r = P_a +P_b \equiv - m \pmod{N}$.   The eigenvalue (\req(LamBBS)) of the Bethe state $\Psi (t_1, \ldots , t_m )$ associated to the Bethe solution $t_i$'s of (\req(BeBBS)) is characterized by the relation
$$
\frac{\omega^m (1-t^N)^L \Lambda (t; t_1, \ldots, t_m)}{h(\omega^{-1} t )^L h(t)^L }F(t) = \omega^m H_{\rm CP}(t )^L  F( \omega^{-1} t) +  H_{\rm CP}(\omega^{-1} t)^L   F(\omega  t ). 
$$
The above relation is the same as (\req(TFfus)) where the factor $\omega^{-P_b}$ in formula (\req(CPTj)) is identified with  $\omega^m$. 
By this, we arrive the conclusion that the $\tau^{(2)}$-eigenspace for the Bethe state $\Psi (t_1, \ldots , t_m )$ satisfies the constrain $-m \equiv P_a +P_b \equiv P_b \pmod{N}$, hence lies in sectors with $P_a  \equiv 0, -m  \equiv P_b$, hence $  - Q \equiv L + m  \pmod{N}$. In particular, the pseudo-vacuum $\Omega_L$ is the ground-state for $P_a = P_b = 0$ with $Q \equiv -L  \pmod{N}$.

We may discuss the algebraic Bethe ansatz of superintegrable CPM in the setting described in the remark of Proposition \ref{prop:BBSBe} with the pseudo-vacuum $\Omega'_L$ and Bethe state $\Psi' (t_1, \ldots , t_m )$ . The Bethe equation (\req(BeBBS')) now becomes (\req(BeBBS)) with $r \equiv -(L+m) \pmod{N}$. 
The eigenvalue (\req(Lam')) for $\Psi'_L (t_1, \ldots , t_m )$ is expressed by 
$$
\frac{(1-t^N)^L \Lambda' (t; t_1, \ldots, t_m)}{h(\omega^{-1} t)^L h(t)^L} F(t)
=  \omega^{L+m} H_{\rm CP} ( t)^L F( \omega ^{-1} t)  + H_{\rm CP}( \omega^{-1}t)^L F( \omega t  ).
$$
Then the $\tau^{(2)}$-eigenspace generated by $\Psi' (t_1, \ldots , t_m )$ has the constraint, $P_b \equiv 0 , \ \ - P_a \equiv L + m  \pmod{N}$, hence $  Q \equiv m  \pmod{N}$. Now the pseudo-vacuum $\Omega'_L$ is the ground-state for $P_b = 0, \ P_a \equiv -L \pmod{N}$ with $Q =0$.

The above discussion about the superintegrable CPM has indicated that the $\tau^{(2)}$-eigenstates in the algebraic-Bethe-ansatz approach appear only in the sectors with $P_a = 0$ or $P_b=0$, not for the rest sectors. The full spectrum of $\tau^{(2)}$-eigenvalues and the symmetry of its degeneracy has been well studied in \cite{R05o} through the chiral Potts transfer matrix as the Baxter's $Q$-operator for the $\tau^{(2)}$-matrix, and the complete detailed structure was obtained by the functional-relation-method. Nevertheless, the algebraic-Bethe-ansatz approach does provide a mechanism to understand some $\tau^{(2)}$- eigenvectors in certain sectors.    

\section{The XXZ Spin Chain of Higher Spin at Roots of Unity \label{sec:6v}}
\setcounter{equation}{0}
In this section we study the fusion relation and Bethe ansatz of XXZ chain of spin $\frac{d-1}{2}$ at roots of unity  for positive integers $d \geq 2$ with the anisotropy parameter $q$. Here the discussion of the  XXZ spin chain  will always assume with 
the {\it even} chain-size $L$.  
We shall denote $\omega = q^2$; when $q$ is a $N$th root of unity, we consider only the {\it odd} $N$ case, then take $q$, $\omega (:= q^2)$, $q^{\frac{1}{2}}$ all to be primitive $N$th roots of unity.

The $L$-operator of the XXZ chain of spin-$\frac{1}{2}$ is the operator with $\CZ^2$-auxiliary and $\CZ^2$-quantum space: 
\be
L (s)  =  \left( \begin{array}{cc}
        L_{0,0}  & L_{0,1}  \\
        L_{1,0} & L_{1,1} 
\end{array} \right) (s) , \ \ s \in \CZ  , 
\ele(6VL)
where the $\CZ^2$-(quantum-space) operator-entries $L_{i,j}$ are  given by 
$$
\begin{array}{l}
L_{0,0} = \left( \begin{array}{cc}
        a  & 0\\
        0& b 
\end{array} \right), \ L_{0,1} = \left( \begin{array}{cc}
        0  & 0\\
        c& 0
\end{array} \right) , \ L_{1,0} = \left( \begin{array}{cc}
        0  & c\\
        0& 0 
\end{array} \right) , \ L_{1,1} = \left( \begin{array}{cc}
        b  & 0\\
        0& a 
\end{array} \right), 
\end{array}
$$
with $a, b, c$ the $q$-dependent $s$-functions
\be
a=a(s)= s q^{\frac{-1 }{2}} - s^{-1} q^{\frac{1 }{2}}, \ \ b= b(s) (= a(qs)) = s q^{\frac{1 }{2}} - s^{-1} q^{\frac{-1 }{2}}, \ \ c= q-q^{-1}.
\ele(6abc)
The operator (\req(6VL)) satisfies the YB relation,
\be
R_{\rm 6v}(s/s') (L(s) \bigotimes_{aux}1) ( 1
\bigotimes_{aux} L(s')) = (1
\bigotimes_{aux} L(s'))( L(s)
\bigotimes_{aux} 1) R_{\rm 6v}(s/s').
\ele(6YB)
where $R_{\rm 6v}$ is the $R$-matrix 
$$
R_{\rm 6v} (s) = \left( \begin{array}{cccc}
        s^{-1} q - s q^{-1}  & 0 & 0 & 0 \\
        0 &s^{-1} - s & q-q^{-1} &  0 \\ 
        0 & q-q^{-1} &s^{-1} - s & 0 \\
     0 & 0 &0 & s^{-1} q - s q^{-1} 
\end{array} \right).
$$
Hence the monodromy matrix of size $L$,
\be
\bigotimes_{\ell=1}^L L_\ell (s) = \left( \begin{array}{cc}
        {\sf A}(s)  & {\sf B}(s) \\
        {\sf C}(s) &  {\sf D}(s) 
\end{array} \right) 
\ele(6mon)
again satisfies the YB relation (\req(6YB)), i.e. the entries ${\sf A, B, C, D}$ form the well known ABCD-algebra, in which the following relations hold: 
\bea(lr)
[{\sf A}(s), {\sf A}(s')]= [{\sf B}(s), {\sf B}(s')]=[{\sf C}(s), {\sf C}(s')]=[{\sf D}(s), {\sf D}(s')]=0 ;& \\ 
{\sf A}(s){\sf B}(s') = f_{s, s'}  {\sf B}(s'){\sf A}(s) -  g_{s, s'} {\sf B}(s){\sf A}(s') , & {\sf A} \leftrightarrow {\sf B}; \ \ {\sf A, B} \longrightarrow {\sf C, D} ; \\
{\sf D}(s){\sf B}(s') = f_{s',s} {\sf B}(s'){\sf D}(s) - g_{s', s} {\sf B}(s){\sf D}(s'), &{\sf B} \leftrightarrow {\sf D}; \ \ {\sf B, D}  \longrightarrow {\sf A, C}  . 
\elea(6ABCD)
where the functions $f_{s, s'}, g_{s, s'}$ are defined by 
\be
f_{s, s'} =  \frac{s^2 q^2 - s'^2  }{q (s^2- s'^2 )} , \ \ g_{s, s'} = \frac{s s'(q^2 - 1)}{q (s^2- s'^2 )}.
\ele(fg)
Then follow the relations
\bea(ll)
{\sf A}(s) \prod_{i=1}^m {\sf B}(s_i) 
&= (\prod_{i=1}^m f_{s, i})  (\prod_{i=1}^m {\sf B}(s_i)) {\sf A}(s) - \sum_{k=1}^m g_{s, k} (\prod_{i=1, i \neq k }^m f_{k, i})  {\sf B} (s)  \prod_{i=1, i \neq k }^m {\sf B}(s_i) {\sf A}(s_k), \\
{\sf D}(s) \prod_{i=1}^m {\sf B}(s_i) 
&= (\prod_{i=1}^m f_{i, s} ) (\prod_{i=1}^m {\sf B}(s_i)) {\sf D}(s) - \sum_{k=1}^m g_{k, s}  (\prod_{i=1, i\neq k}^m f_{i, k} ){\sf B}(s) \prod_{i=1, i \neq k }^m {\sf B}(s_i) {\sf D}(s_k)  ; \\
&{\sf A}(s) \leftrightarrow {\sf D}(s) , {\sf B}(s) \leftrightarrow {\sf C}(s) .
\elea(6ADs)
Here we write only $i$ for $s_i$ in the subscripts of $f_{s ,s'}, g_{s, s'}$.

As the matrix $R_{\rm 6v} (q)$ is of rank 1, the quantum determinant of the YB solution $L(s)$ in (\req(6YB)) is defined by 
\be
R_{\rm 6v} (q) (L(q s) \bigotimes_{aux}1) ( 1
\bigotimes_{aux} L(s)) = (1 \bigotimes_{aux} L(s))( L(qs)
\bigotimes_{aux} 1) R_{\rm 6v} (q) =: {\rm det}_q L(s) \cdot  R_{\rm 6v} (q),
\ele(6qdet)
or equivalently, 
\bea(rl)
&{\sf A} (qs) {\sf C} (s) = {\sf C} (qs) {\sf A} (s) ,  ~ \ ~ {\sf B} (qs) {\sf D} (s)  = {\sf D} (qs) {\sf B} (s) , \\
& {\sf B} (s) {\sf A} (qs) = {\sf A} (s) {\sf B} (qs) , ~ \ ~ {\sf C} (s) {\sf D} (qs)  = {\sf D} (s) {\sf C} (qs) ; \\
{\rm det}_q L(s) =& {\sf A} (qs) {\sf D} (s) - {\sf C} (qs) {\sf B} (s) = {\sf D} (qs) {\sf A} (s) - {\sf B} (qs) {\sf C} (s) \\
= & {\sf D} (s) {\sf A} (qs) - {\sf C} (s) {\sf B} (qs) = {\sf A} (s) {\sf D} (qs)  - {\sf B} (s) {\sf C} (qs).
\elea(6qdf)
For the local $L$-operator (\req(6VL)), ${\rm det}_q L(s) = a( s) a(q^2 s)$, so the quantum determinant of (\req(6mon)) is equal to $ a(s)^L a(q^2 s)^L$. The statements in Lemma \ref{lem:cd} are valid for the operator $L(s)$ by replacing ${\tt L}( \omega t) \otimes_{aux} {\tt L}(t), {\rm det}_q {\tt L}(t)$ by $L( s) \otimes_{aux} L (qs), {\rm det}_q L(s)$ in (\req(qCom)), and ${\tt L}( \omega^{j-2}t) \otimes_{aux} \cdots  \otimes_{aux} {\tt L}(t)$ by $L( s) \otimes_{aux} \cdots  \otimes_{aux} L(q^{j-2}s)$
in (\req(gdet)). Now we consider the transfer matrix of the six-vertex model of even size $L$, defined by the trace of (\req(6mon)). Denote
$$
T^{(2)} (t) = s^{2L} ({\sf A}(s) + {\sf D}(s) ), \ \ t:= qs^2 \ \in \CZ.
$$  
Then $T^{(2)} (t)$ form a family of $t$-polynomial operators acting on the quantum space $\stackrel{L}{\otimes} \CZ^2$. As in the discussion of subsection \ref{ssec.FBBS}, we constructed in \cite{R06Q}  the fusion matrices $T^{(j)} (t)$ $(0 \leq j \in \ZZ)$ from the fused $L$-operator $L^{(j)}(s)$ associated to $L(s)$ in (\req(6VL)), which is a matrix with $\CZ^2$-quantum and $\CZ^j$-auxiliary space, as follows. With the basis in (\req(Cjb)) for the $\CZ^j$-auxiliary space, 
$L^{(j)}(s) = \bigg( L^{(j)}_{k, l} (s)\bigg)_{0 \leq k, l \leq j-1}$ where 
the $\CZ^2$-(quantum-space) operator $L^{(j)}_{k, l}(s)$ is defined by\footnote{One may use $L^{(j)}$-operator defined by revising the order of $L(s) \otimes_{aux} L(qs)\otimes_{aux} \cdots \otimes_{aux} L(q^{j-2}s)$ in its definition, replaced by  $L(q^{j-2}s) \otimes_{aux} L(q^{j-3}s)\otimes_{aux} \cdots \otimes_{aux} L(s)$ as the order in (\req(Gj)). The latter form is related to the first expression in (\req(6qdet)), instead of the second one in the former case.}
\be
L^{(j)}_{k, l} (s) = \frac{\langle e^{(j)*}_k | L(s) \otimes_{aux} L(qs)\otimes_{aux} \cdots \otimes_{aux} L(q^{j-2}s) | e^{(j)}_l \rangle}{ \prod_{i =0}^{j-3} b( q^i s) } .
\ele(Lj)
The relations, (\req(Gj)) (\req(fzero)) (\req(fj-1)), now turn to  
\bea(l)
L^{(j+1)}_{k, l}(s) = \frac{1}{b( q^{j-2} s)} \langle e^{(j+1)*}_k | L^{(j)}(s) \otimes_{aux} L(q^{j-1}s) | e^{(j+1)}_l \rangle ; \\
\langle e^{(j+1)*}_l | L^{(j)}(s) \otimes_{aux} L(q^{j-1}s) | f^{(j-1)}_k \rangle = 0 ; \\
\langle f^{(j-1)*}_k | L^{(j)}(s) \otimes_{aux} L(q^{j-1}s) | f^{(j-1)}_l \rangle = b(q^{j-1}s) \langle e^{(j-1)*}_k | L^{(j-1)}(s)  | e^{(j-1)}_l \rangle .
\elea(6Ljj)
From the first relation in above, one obtains the explicit form of $L^{(j)}_{k, l}(s)$ for $j \geq 2$ by induction argument (\cite{R06Q} Proposition 4.2): $L^{(j)}_{k, l} (=L^{(j)}_{k, l}(s)) \ (0 \leq k, l \leq j-1)$  are zeros except $k- l = 0, \pm 1$, in which cases 
$$
L^{(j)}_{k, k} = \left( \begin{array}{cc}
     a(q^k s) &0\\
     0 & a( q^{j-1-k} s)   
\end{array} \right), \ \ L^{(j)}_{k+1, k} = \left( \begin{array}{cc}
     0 & q^{j-1-k} -q^{-j+1+k} \\
     0 & 0  
\end{array} \right), \ \ L^{(j)}_{k-1, k} = \left( \begin{array}{cc}
     0 & 0 \\
     q^k -q^{-k} & 0  
\end{array} \right).  
$$
The fusion matrix $T^{(j)} (t)$ is defined by $T^{(0)} = 0$, $T^{(1)}(t) = (1- \omega^{-1} t)^L$, and for $j \geq 2$,    
\be
T^{(j)} (t) = (sq^{\frac{j-2}{2}})^L {\rm tr}_{aux} ( \bigotimes_{\ell=1}^L L^{(j)}_\ell (s)), \ \ t = q s^2  \in \CZ . 
\ele(TjL)
The relations in (\req(6Ljj)) and the explicit form of $L^{(j)}_{k, l}$'s  in turn yield the fusion relation of $T^{(j)}$'s with the truncation identity:
\bea(cl) 
T^{(j)}( t) T^{(2)}(\omega^{j-1} t) =   (1-\omega^{j-1} t)^L T^{(j-1)}(t)  + (1-\omega^{j-2}t)^L T^{(j+1)}(t),  & j \geq 1 , \\
T^{(N+1)}(t) =   T^{(N-1)}( \omega t) + 2 ( 1- \omega^{-1} t)^L , &
\elea(6TfusA)
parallel to the fusion relations, (\req(fus)) and  (\ref{fusB}), in the generalized $\tau^{(2)}$-model.

\subsection{Fusion relation and algebraic Bethe ansatz in XXZ spin chain of  higher spin  \label{ssec.6FuBe}}
When interchanging the auxiliary and quantum spaces of the $d$th fused $L$-operator $L^{(d)}(s)$ for a positive integer $d$, we arrive the $L$-operator of XXZ chain of spin-$\frac{d-1}{2}$, which is the matrix with the $\CZ^2$-auxiliary and $\CZ^d$-quantum space: 
\be
{\sf L} (s)  =  \left( \begin{array}{cc}
        {\sf L}_{0,0}  & {\sf L}_{0,1}  \\
        {\sf L}_{1,0} & {\sf L}_{1,1} 
\end{array} \right) (s) , \ \ s \in \CZ  , 
\ele(s6VL)
where the entries $({\sf L}_{a,b})^{i'}_i \ (0 \leq i, i' \leq d-1)$ of ${\sf L}_{a,b}$ are zeros except $({\sf L}_{0,0})^{i}_i = ({\sf L}_{1,1})^{d-i-1}_{d-i-1} = a ( q^i s )$, $( 0 \leq i \leq d-1)$, and $({\sf L}_{0,1})^{i}_{i+1} = q^{d-1-i} - q^{-d+1+i}$,  $ 
({\sf L}_{1,0})^{i+1}_i = q^{i+1} - q^{-i-1}$, $(0 \leq i < d-1)$.
Here ${\sf L}_{a,b}$ are operators of the quantum space $\CZ^d$ with the standard basis ${\sf e}^i ~  (i=0, \ldots, d-1)$.
It is well-known that the above expression of ${\sf L}_{a,b}$ gives the $d$-dimensional irreducible representation of $U_q(sl_2)$ (see, e.g., \cite{KiR, KRS}),
\bea(ll)
{\sf L}_{0,0} = sq^{\frac{d-2}{2}} \widehat{K}^{\frac{-1}{2}} - s^{-1}q^{\frac{-(d-2)}{2}} \widehat{K}^{\frac{1}{2}} , & {\sf L}_{0,1} = (q-q^{-1}) \hat{e}_-, \\ 
{\sf L}_{1,0} = (q-q^{-1}) \hat{e}_+ , & {\sf L}_{1,1} = sq^{\frac{d-2}{2}}\widehat{K}^{\frac{1}{2}} - s^{-1}q^{\frac{-(d-2)}{2}} \widehat{K}^{\frac{-1}{2}} ,
\elea(Uq6d)
with the relations, $\widehat{K} \hat{e}_\pm \widehat{K}^{-1} = q^{\pm 2} \hat{e}_\pm$ and $[\hat{e}_+, \hat{e}_- ] = \frac{\widehat{K} - \widehat{K}^{-1}}{q-q^{-1}}$, where $\widehat{K}^{\frac{1}{2}} = q^{S^z}$ and $S^z$ the spin-operator of $\CZ^d$: $S^z  ={\rm dia}[\frac{d-1}{2}, \frac{d-3}{2}, \ldots, \frac{-d+1}{2}]$.
By the direct verification, one finds the $L$-operator (\req(s6VL)) again satisfies the six-vertex YB relation (\req(6YB)), hence the $L$th monodromy matrix 
\be
\bigotimes_{\ell=1}^L  {\sf L}_\ell (s) = {\sf L}_1(s) \otimes \cdots \otimes {\sf L}_L (s) =  \left( \begin{array}{cc} {\sf A}(s)  & {\sf B} (s) \\
      {\sf C}(s) & {\sf D}(s)
\end{array} \right), \ \ {\sf L}_\ell (t):= {\sf L}(t) \ {\rm at \ site} \ \ell,
\ele(mM6d)
has the entries forming 
 the ABCD-algebra with the pseudo-vacuum $\Omega $ and quantum determinant (\req(6qdf)): 
\be
\Omega  = \stackrel{L}{\otimes} {\sf e}^0 \ \in \stackrel{L}{\otimes} \CZ^d , \ \ \ {\rm det}_q {\sf L}_L (s) = a(q^d s )^L a (s)^L \cdot {\rm id}  .
\ele(6dpv)
The trace of (\req(mM6d)) defines a family of commuting $(\stackrel{L}{\otimes}\CZ^d)$-operators, called the transfer matrix of the XXZ chain of spin-$\frac{d-1}{2}$:
$$
{\sf t}(s) = {\sf A}(s) + {\sf D}(s), 
$$
which commute with $q^{S^z}=  \stackrel{L}{\otimes} \widehat{K}^{\frac{1}{2}}$, where $S^z$ is the spin-operator of $\stackrel{L}{\otimes} \CZ^d$. 
Similar to the discussion of (\req(Gj)) and (\req(Lj)), we consider the fused $L$-operator associated with (\req(s6VL))\footnote{Note that for $d > 2$, ${\sf L}^{(j)}_{k, l} (s)$ have no common factors in contrast to the case $d=2$ in (\req(Lj)). Hence we use the notation ${\sf L}$ for the $L$-operator for a general $d$ to distinguish it from the $L$-operator (\req(Lj)) for $d=2$, in which case ${\sf L}^{(j)}_{k, l} (s) = (\prod_{i =0}^{j-3} b( q^i s)) L^{(j)}_{k, l} (s)$. However the fusion matrix ${\sf t}^{(j)}$ will still be divisible by certain factors after multiplying the normalization factors in (\req(6Tjt)) and (\req(6dTj)). }: 
$$
{\sf L}^{(j)}_{k, l} (s) = \langle e^{(j)*}_k | {\sf L}(s) \otimes_{aux} {\sf L}(qs)\otimes_{aux} \cdots \otimes_{aux} {\sf L}(q^{j-2}s) | e^{(j)}_l \rangle ,
$$
and define the fusion matrix
$$
{\sf t}^{(j)} (s) = {\rm tr}_{\CZ^j} (\bigotimes_{\ell=1}^L  {\sf L}^{(j)}_\ell (s)),
$$
with ${\sf t}^{(2)} (s)= {\sf t} (s)$.
Then ${\sf t}^{(j)} (s)$  form a family of commuting operators of $\stackrel{L}{\otimes} \CZ^d$.
Similar to (\req(6Ljj)), we now have
$$
\begin{array}{l}
{\sf L}^{(j+1)}_{k, l}(s)  =  \langle e^{(j+1)*}_k | {\sf L}^{(j)}(s) \otimes_{aux} {\sf L} (q^{j-1}s) | e^{(j+1)}_l \rangle, ~ ~ 
\langle e^{(j+1)*}_l | {\sf L}^{(j)}(s) \otimes_{aux} {\sf L}(q^{j-1}s) | f^{(j-1)}_k \rangle = 0 , \\
\langle f^{(j-1)*}_k | {\sf L}^{(j)}(s) \otimes_{aux} {\sf L} (q^{j-1}s) | f^{(j-1)}_l \rangle = {\rm det}_q {\sf L}(q^{j-2}s) \langle e^{(j-1)*}_k | {\sf L}^{(j-1)}(s)  | e^{(j-1)}_l \rangle ,
\end{array}
$$
then follows the fusion relation of the XXZ chain of spin-$\frac{d-1}{2}$, parallel to  (\req(fus)) in the generalized $\tau^{(2)}$-model, by setting ${\sf t}^{(0)}=0, {\sf t}^{(1)}= I$:
\bea(l)
{\sf t}^{(j)} (s) {\sf t}^{(2)}(q^{j-1} s) =  a(q^{d+j-2} s )^L a(q^{j-2} s)^L {\sf t}^{(j-1)}(s)  + {\sf t}^{(j+1)}(s) , \ \ j \geq 1.
\elea(6dfus)
Using the $t$-variable  in (\req(TjL)), and normalizing ${\sf t}^{(j)} (s)$ by
\be
{\cal T}^{(j)} (t) = (s^{(j-1)} q^{\frac{(j-1)(d+j-4) }{2}})^L  {\sf t}^{(j)} (s) , \ \  t= qs^2 , 
\ele(6Tjt)
one can write the fusion relation (\req(6dfus)) in terms of $t$-polynomial operators ${\cal T}^{(j)} (t)$:
\be
{\cal T}^{(j)} (t) {\cal T}^{(2)}(\omega^{j-1} t) 
=  h(\omega^{j-3}t)^L h(\omega^{d+j-3}t)^L {\cal T}^{(j-1)}(t)  + {\cal T}^{(j+1)}(t) \ \ (j \geq 1)
\ele(6Fud)
with ${\cal T}^{(0)}=0, {\cal T}^{(1)}= I$, and $h(t)$ in (\req(hfun)).

We now discuss the algebraic Bethe ansatz of XXZ chain of spin-$\frac{d-1}{2}$ \cite{TakF}. With the pseudo-vacuum $\Omega$ in (\req(6dpv)), $
{\sf A}(s)  \Omega = a(s)^L \Omega$ , $
{\sf D}(s)  \Omega = a(q^{d-1} s)^L \Omega$ , and $ {\sf C}(s)  \Omega = 0 $. 
For $m$ square-distinct non-zero complex numbers $s_j, 1 \leq j \leq m$, we consider the vector $\Phi (s_1, \ldots , s_m )= \prod_{i=1}^m{\sf B} (s_i)\Omega \in \stackrel{L}{\otimes} \CZ^d$ with  $S^z = \frac{L(d-1)-2m}{2}$, 
and define the $s$-function  
\be
\Lambda (s; s_1, \ldots, s_m) = a(s)^L \prod_{i=1}^m f_{s, i} +  a(q^{d-1} s)^L \prod_{i=1}^m f_{i, s} ,
\ele(6Lam)
where $f_{s,s'}$ are functions in (\req(fg)). The regular-criterion for the $s$-function (\req(6Lam)) in the non-zero $s$-domain  is given by the Bethe equation
\be
\frac{ a(s_i )^L }{a(q^{d-1} s_i )^L} = -  q^{2m} \prod_{k =1}^m \frac{ s^2_k  - q^{-2} s^2_i }{  s^2_k - q^2  s^2_i  } , \ \ i =1, \ldots, m ,
\ele(6dBes)
in which case by (\req(6ADs)) and the same argument in Proposition \ref{prop:BBSBe}, $\Phi (s_1, \ldots , s_m )$ is an eigenvector of the transfer matrices ${\sf t}(s)$ with the eigenvalue $\Lambda (s; s_1, \ldots, s_m)$. Using the variable $t$ in (\req(6Tjt)) and $t_i = qs_i^2$, (\req(6dBes)) and (\req(6Lam)) now take the form
\begin{eqnarray}
\frac{h(\omega^{-1} t_i )^L }{h(\omega^{d-2} t_i )^L}  = - \omega^{- \frac{L(d-1)-2m}{2}}  \frac{F(\omega^{-1}t_i )}{ F( \omega t_i)} \ ~ ~ \ (i=1, \ldots, m) , \ \ F(t):= \prod_{i=1}^m (1- t_i^{-1}t) ; \label{6dBe} \\
s^L q^{\frac{L(d-2)}{2}} \Lambda (s; s_1, \ldots, s_m) = h(\omega^{d-2}t)^L  \frac{F(\omega^{-1}t )}{ F(t)} + \omega^{\frac{L(d-1)-2m}{2}} h( \omega^{-1}t)^L   \frac{ F( \omega t)}{F(t) } . \label{6dLm}
\end{eqnarray}

In the root-of-unity case where both $q, \omega$ are $N$th roots of unity, we are going to derive the evaluation polynomial for the root-of-unity symmetry study of the XXZ chain of spin-$\frac{d-1}{2}$ with $2 \leq d \leq N$. Define the $t$-polynomial
\be
H (t) \ (= H_{d} (t)) : = \frac{1-t^N}{\prod_{k=d-1}^{N-1} h(\omega^k t)} = \prod_{k=0}^{d-2} (1- \omega^k t), \ \ \ \ \ 2 \leq d \leq N.
\ele(H6V)
Using the above function $H(t)$, one writes the Bethe equation (\ref{6dBe}) in a similar form as (\req(BeBBS)):
\be
\frac{H (\omega^{-1} t_i )^L }{H (t_i )^L}  = - \omega^{- r}  \frac{F(\omega^{-1}t_i )}{ F( \omega t_i)} \ ~ ~ \ (i=1, \ldots, m) , \ \ F(t):= \prod_{i=1}^m (1- t_i^{-1}t) ,
\ele(6dBet)
where $0 \leq r \leq N-1$, $ r \equiv \frac{L(d-1)-2m}{2} \ (= S^z) \pmod{N}$. Now we consider only those Bethe states $\Phi (s_1, \ldots , s_m )$ so that $\{ t_1, \cdots, t_m \} $ contains no complete $N$-cyclic string, (which means $\{ \omega^j t_0 \}_{j \in \ZZ_N}$ for some $t_0 \neq 0$). Normalize the ${\cal T}^{(j)}$-operator (\req(6Tjt)) by\footnote{For $d=2$, the ${\sf T}^{(j)}$ here differs from $T^{(j)}$ in (\req(TjL)) only by the factor $\omega^{(j-1)S^z}$: ${\sf T}^{(j)}(t) = \omega^{-(j-1)S^z} T^{(j)}(t)$, hence in sectors $S^z \equiv 0 \pmod{N}$, it defines the same operator which was used in the discussion of \cite{R06Q}.} 
\be
 {\sf T}^{(j)} (t) := \frac{\omega^{-(j-1)S^z} (1-t^N) {\cal T}^{(j)}(t)}{\prod_{k=d-2}^{N+j-3} h(\omega^k t) } , \ \ j \geq 1 . 
\ele(6dTj)
By (\ref{6dLm}), the ${\sf T}^{(2)}$-eigenvalue $\lambda^{(2)} (t; t_1, \ldots, t_m)$ 
for the Bethe state $\Phi (s_1, \ldots , s_m )$ is the $t$-polynomial characterized by the relation
\be 
\lambda^{(2)} (t; t_1, \ldots, t_m) F(t) = \omega^{-r} H (t)^L  F(\omega^{-1}t )  + H ( \omega^{-1}t)^L F( \omega t) .
\ele(6lam)
The relation (\req(6Fud)) in turn yields  the fusion relation of ${\sf T}^{(j)}$'s with ${\sf T}^{(0)}=0$, ${\sf T}^{(1)} (t) = H (\omega^{-1} t)^L$ and 
\be
{\sf T}^{(j)}(t) {\sf T}^{(2)}(\omega^{j-1} t) =  \omega^{-S^z} H(\omega^{j-1} t)^L {\sf T}^{(j-1)}(t)  + H(\omega^{j-2}t)^L {\sf T}^{(j+1)}(t),  \  j \geq 1 .
\ele(6dfut)
Using the same argument as \cite{R06Q}, (where formulas (3.8) and (3.9) correspond  to (\req(6lam)) and (\req(6dfut)) here), one obtains the expression of ${\sf T}^{(j)}$-eigenvalue $\lambda^{(j)} (t; t_1, \ldots, t_m)$ for the Bethe state $\Phi (s_1, \ldots , s_m )$: 
\be 
\lambda^{(j)} (t; t_1, \ldots, t_m) = F( \omega^{-1} t) F( \omega^{j-1}t) \sum_{k=0}^{j-1} \frac{ H (\omega^{k-1} t )^L \omega^{-kr} }{F(\omega^{k-1}t)F(\omega^k t)} \ , \ \ \ \ j \geq 1 ,
\ele(lamj)
which is a $t$-polynomial by the Bethe equation (\req(6dBet)). Then one arrives the boundary fusion relation for ${\sf T}^{(j)}$'s ( on the space spanned by all eigenspaces of the Bethe states): 
\be
{\sf T}^{(N+1)}(t) = \omega^{-S^z} {\sf T}^{(N-1)}( \omega t)
  + 2 H ( \omega^{-1} t)^L . 
\ele(6fBdy)
Formulas (\req(6dfut)) and (\req(6fBdy)) are the complete fusion relations of the XXZ spin chain of spin-$\frac{d-1}{2}$ at roots of unity, which in $d=2$ case, are equivalent to (\req(6TfusA)).
By (\req(lamj)), one writes the ${\sf T}^{(N)}$-eigenvalue in the form
$$ 
\lambda^{(N)} (t; t_1, \ldots, t_m) =  \omega^{-r}t^r F( \omega^{-1} t)^2  P_{\rm 6V} (t^N),
$$
where $P_{\rm 6V}( \xi)$ is the function defined by 
\be 
P_{\rm 6V} (t^N):= \sum_{k=0}^{N-1} \frac{ H (\omega^k t)^L (\omega^k t )^{-r}}{F(\omega^k t)F(\omega^{k+1} t)},  \ \ F(t) := \prod_{i=1}^m (1- t_i^{-1} t).
\ele(p6V)
The Bethe relation (\req(6dBet)) for $t_i$'s is the polynomial-condition for $P_{\rm 6V}$, in which case by (\req(lamj)), the ${\sf T}^{(j)}$-eigenpolynomial and $P_{\rm 6V}$-polynomial are related by 
$$
\begin{array}{cl}
{\sf T}^{(j)}(t) + \omega^{-jr} {\sf T}^{(N-j)}(\omega^j t) = (\omega^{-1} t)^r F( \omega^{-1}t) F( \omega^{j-1}t) P_{\rm 6V}(t^N ), & 0 \leq j \leq N , \\
{\sf T}^{(N)}(t) = (\omega^{-1} t)^r F( \omega^{-1} t)^2  P_{\rm 6V}(t^N ) .& 
\end{array}
$$
The above relations reflect the $QQ$-relations in the functional-relation setting of  XXZ spin chain of higher spin for a proper $Q$-operator which encodes the root-of-unity property; the construction of such $Q$-operator is not known yet except the spin-$\frac{1}{2}$ case \cite{R06Q}. 
Note that for $d=2, N$, $H (t)= 1-t$ , $\frac{1-t^N}{1- \omega^{-1} t}$, and $P_{\rm 6V}(\xi)$  is the Drinfeld polynomial in \cite{De05, FM01, R06Q, NiD} for the root-of-unity XXZ chain of spin-$\frac{1}{2}$, $\frac{N-1}{2}$ respectively. For a general $d$, 
the $P_{\rm 6V}$-polynomial is indeed the evaluation polynomial for the root-of-unity symmetry of XXZ spin chain of higher spin, which will be verified in the next subsection.

\subsection{The $sl_2$-loop-algebra symmetry of in the root-of-unity XXZ spin chain of higher spin  \label{ssec.6symd}}
In this subsection, we are going to show the $sl_2$-loop-algebra symmetry of the XXZ chain of spin-$\frac{d-1}{2}$ with the $N$th root-of-unity anisotropic parameter $q$, {\it even} chain-size $L$, odd $N$, and the total spin $S^z \equiv 0 \pmod{N}$ for $ 2 \leq d \leq N$. 
The $P_{\rm 6V}$-polynomial in (\req(p6V)) will be shown as the evaluation polynomial, i.e., the Drinfeld polynomial, for the $sl_2$-loop-algebra representation. For $d=2$, the result is known by works in \cite{DFM, FM01, De05} , and for $d=N$, the conclusion is obtained in \cite{NiD}. Here we are going to derive the root-of-unity symmetry of XXZ spin chain of higher spin along the line in \cite{DFM, FM01}.

Define
$$
S^\pm =  \sum_{i=1}^L  \underbrace{\widehat{K}^{\frac{1}{2}} \otimes \cdots \otimes \widehat{K}^{\frac{1}{2}}}_{i-1}\otimes \hat{e}_\pm \otimes  \underbrace{\widehat{K}^{\frac{-1}{2}} \otimes \cdots \otimes \widehat{K}^{\frac{-1}{2}}}_{L-i} , \ \
T^\pm  =  \sum_{i=1}^L  \underbrace{\widehat{K}^{\frac{-1}{2}} \otimes \cdots \otimes \widehat{K}^{\frac{-1}{2}}}_{i-1}\otimes \hat{e}_\pm \otimes  \underbrace{\widehat{K}^{\frac{1}{2}} \otimes \cdots \otimes \widehat{K}^{\frac{1}{2}}}_{L-i}. 
$$
The leading and lowest terms of entries of (\req(mM6d)) are given by 
$$
\begin{array}{ll}
{\sf A}_\pm = \lim_{s^{\pm} \rightarrow \infty} (\pm s)^{\mp L}q^{\frac{\mp L(d-2)}{2}} {\sf A}(s) , & {\sf B}_\pm = \lim_{s^{\pm} \rightarrow \infty} (\pm s)^{\mp (L-1)}q^{\frac{\mp (L-1)(d-2)}{2}} \frac{{\sf B}(s)}{q-q^{-1}} ,  \\ 
{\sf C}_\pm = \lim_{s^{\pm} \rightarrow \infty} (\pm s)^{\mp (L-1)}q^{\frac{\mp (L-1)(d-2)}{2}} \frac{{\sf C}(s)}{q-q^{-1}}, & {\sf D}_\pm = \lim_{s^{\pm} \rightarrow \infty} (\pm s)^{\mp L}q^{\frac{\mp L(d-2)}{2}} {\sf D}(s).
\end{array}  
$$
Using (\req(Uq6d)), one finds
\be
q^{ \pm S^z}= {\sf A}_\mp = {\sf D}_\pm , \ ~ \ T^- = {\sf B}_+ , \  S^- = {\sf B}_- , \ ~ \ S^+ = {\sf C}_+ , \ T^+ = {\sf C}_- ,
\ele(Uslq)
which give rise to the representation of $U_q(\widehat{sl}_2)$ on $\stackrel{L}{\otimes} \CZ^d$ (by $k_0^{-1} = k_1 = q^{2 S^z}$, $e_0 = T^- ,  f_0 = T^+$, $e_1 = S^+ ,   f_1 = S^-$).
Consider the normalized $n$th power of operators $S^\pm, T^\pm$ as in \cite{DFM, NiD},
$S^{\pm (n)} = \frac{S^{\pm n}}{[n]!}$ , $T^{\pm (n)} = \frac{T^{\pm n}}{[n]!}, \ (n \geq 0)$ for a generic $q$, then take the limit as $q$ being the $N$th root of unity, where $[n] = \frac{q^n-q^{-n}}{q-q^{-1}}$ , $[n]!= \prod_{i=1}^n [i]$ and $[0]!:=1$. For $q^N=1$, one obtains the operators,
$$
\begin{array}{ll}
S^{\pm (N)} =& \sum_{ 0 \leq k_i < d, \  k_1+\cdots+ k_L=N } \frac{1}{[k_1]! \cdots [k_L]!} \otimes_{i=1}^L  \widehat{K_i}^{\frac{-1}{2}( \sum_{j<i} - \sum_{j>i})k_j  } \hat{e}_{i  \pm }^{k_i}, \\
T^{\pm (N)} =& \sum_{ 0 \leq k_i < d, \  k_1+\cdots+ k_L=N } \frac{1}{[k_1]! \cdots [k_L]!} \otimes_{i=1}^L  \widehat{K_i}^{\frac{1}{2}(\sum_{j<i} - \sum_{j>i})k_j  }  \hat{e}_{i  \pm}^{k_i},
\end{array}
$$
with the relation $T^{\pm (N)} = R S^{\mp (N)} R^{-1}$, where $R$ is the spin-inverse operator of $\stackrel{L}{\otimes} \CZ^d$. Note that by $\otimes_{i=1}^L\widehat{K_i}^{\frac{\pm 1}{2}(\sum_{j<i} - \sum_{j>i})k_j} \hat{e}_{i  \pm }^{k_i} = \otimes_{i=1}^L \hat{e}_{i  \pm }^{k_i} \widehat{K_i}^{\frac{\pm 1}{2}(\sum_{j<i} - \sum_{j>i})k_j  }$, the order of powers of $\widehat{K_i}$ and $\hat{e}_{i \pm }$ in the above formulas can be interchanged for the same operator. The $\frac{2S^z}{N}$ and $S^{\pm (N)}, T^{\pm (N)}$ give rise to a $sl_2$-loop-algebra representation with their relations to Chevalley generators as follows:
\be
- H_0 = H_1 = \frac{2S^z}{N} ; \ E_0 = T^{- (N)}, \ E_1 = S^{+ (N)}, \ F_0 = T^{+ (N)},  \ F_1 = S^{- (N)}.
\ele(sl2)
The above Chevalley basis is related to  the mode basis of the $sl_2$-loop algebra, $e(n), f(n), h(n) ~ (n\in \ZZ)$ by  $e(0)= S^{+ (N)}, f(0)= S^{- (N)}$, $e(-1)= T^{+ (N)}, f(1)= T^{- (N)}$, $h(0)= \frac{2S^z}{N} $.

Set $m=N, s_i = x q^{\frac{N+1-2i}{2}} \ (i = 1 , \ldots, N)$ in  (\req(6ADs)), one obtains
$$
\begin{array}{lll} 
{\sf A}(s) \prod_{i=1}^N {\sf B}(s_i) =  \frac{ x^2  -s^2 q^{N+1}}{x^2 q^{N} - s^2 q }  \prod_{i=1}^N {\sf B}(s_i) {\sf A}(s) +  \frac{s s_1 (q^2 - 1)}{q (s_1 ^2- s^2 )} \frac{[N]!}{[N-1]!}  {\sf B} (s)  \prod_{i=2 }^N {\sf B}(s_i) {\sf A}(s_1), & {\sf A, B} \leftrightarrow {\sf D, C} ; \\
{\sf D}(s) \prod_{i=1}^N {\sf B}(s_i)=  \frac{ x^2 -s^2 q^{-N-1} }{x^2 q^{-N} - s^2 q^{-1} }  \prod_{i=1}^N {\sf B}(s_i) {\sf D}(s) -  \frac{s s_N (q^2 -1)}{q (s_N^2- s^2 )}  \frac{[N]!}{[N-1]!} {\sf B}(s) \prod_{i=1}^{N-1} {\sf B}(s_i) {\sf D}(s_N)  , & {\sf D, B} \leftrightarrow {\sf A, C} .
\end{array}
$$
Multiplying each ${\sf B}(s_i), {\sf C}(s_i)$ in the above formulas by the factor $(\pm s)^{\mp (L-1)}_i q^{\frac{\mp (L-1) (d-2)}{2}}(q-q^{-1})^{-1}$ at a generic $q$, then taking the $\infty$-limit for $s_i^{\pm 1}$ at $q^N=1$, one arrives
$$
\begin{array}{lll} 
{\sf A}(s) S^{\pm (N)} &=   S^{\pm (N)} {\sf A}(s) &+( \mp  sq^{\frac{d-2}{2}})^{\pm 1} S^{\pm (N-1)} {}^{{\sf C}(s)}_{{\sf B}(s)}   q^{ \mp S^z},  \\
{\sf D}(s) S^{\pm (N)} &=  S^{\pm (N)} {\sf D}(s) &- (\mp sq^{\frac{d-2}{2}})^{\pm 1} S^{\pm (N-1)}  {}^{{\sf C}(s)}_{{\sf B}(s)}  q^{\pm S^z} , \\
{\sf A}(s) T^{\mp (N)} &=   T^{\mp (N)} {\sf A}(s)  &- (\mp sq^{\frac{d-2}{2}})^{\pm 1}  T^{\mp (N-1)} {}^{{\sf B}(s)}_{{\sf C}(s)} q^{\mp S^z},  \\
{\sf D}(s) T^{\mp (N)} &=  T^{\mp (N)} {\sf D}(s)  & + (\mp sq^{\frac{d-2}{2}})^{\pm 1} T^{\mp (N-1)} {}^{{\sf B}(s)}_{{\sf C}(s)} q^{\pm S^z}, 
\end{array}
$$
which imply 
$$
{ [}{\sf t}(s), S^{\pm (N)}  ]= (sq^{\frac{d-2}{2}})^{ \pm 1} S^{\pm  (N-1)}{}^{{\sf C}(s)}_{{\sf B}(s)} ( q^{S^z} - q^{-S^z}), \ [ {\sf t}(s), T^{\mp  (N)}  ] = -(sq^{\frac{d-2}{2}})^{\pm 1} T^{\mp (N-1)} {}^{{\sf B}(s)}_{{\sf C}(s)} ( q^{S^z} - q^{-S^z}),
$$
hence $S^{\pm (N)}, T^{\pm (N)}$ commute with the transfer matrix ${\sf t}(s)$ in sectors $S^z \equiv 0 \pmod{N}$. Denote by $V$ the subspace of $\stackrel{L}{\otimes} \CZ^j$ consisting all vectors with $S^z \equiv 0 \pmod{N}$.  Therefore the ${\sf T}^{(2)}$- and ${\sf T}^{(j)}$-operators (\req(6dTj)) when restricted on $V$ possess the $sl_2$-loop-algebra symmetry by (\req(sl2)). Hence each degenerate eigenspace gives rise to a finite-dimensional irreducible representation of 
$sl_2$-loop algebra. It is shown in \cite{ChP} that every such representation is obtained by an irreducible representation of $\stackrel{M}{\oplus} sl_2$ on $\otimes_{k=1}^M \CZ^{\delta_k}$ through evaluating the loop-variable at a finite number of distinct non-zero complex numbers, $a_1, \ldots, a_M$, with the evaluation (Drinfeld) polynomial defined by $P(\xi) = \prod_{k=1}^M ( 1- a_k \xi)^{\delta_k - 1}$. Hence the representation space is generated the  
highest weight vector $|v \rangle$, i.e., the unique 
vector (up to scalars) with the highest weight among $\frac{2S^z}{N}$-eigenspaces, and
$P(\xi)$ is expressed by 
$$
P(\xi) = \sum_{r \geq 0} \mu_r (-\xi)^r , \ \ \ \  \frac{S^{+ (N) r} T^{- (N)r}}{(r!)^2}  |v \rangle = \mu_r |v \rangle 
$$ 
with the degree equal to $\sum_{k=1}^M (\delta_k-1)$ (\cite{De05} (1.9), \cite{FM01}(1.17)). The polynomial $P(\xi)$ can be determined by the following current (\cite{FM01} (1.20))\footnote{Here we use the current slightly different from the one used in  \cite{FM01} (1.20) since we employ another, but equivalent, set of representatives for the Chevalley basis in this paper.}
\be
E^- ( \xi ) = \sum_{n=0}^\infty f(n) \xi^n , 
\ele(Ec)
whose pole-structure coincides with the zero-structure of $P(\xi)$. Conjecturally, every Bethe state is annihilated by $S^{+(N)}, T^{+(N)}$ (\cite{FM01} (1.11), \cite{De05} (5.1) (5.2) and Sect. 5.2), by which a Bethe state $\Phi (s_1, \ldots , s_m )$ is the highest weight vector of the $sl_2$-loop-algebra representation to which it belongs. We are going to determine the current (\req(Ec)) by the "vanishing $N$-string" method in  \cite{FM01}.

Consider the average of a commuting family of operators ${\sf O}(s)$:
$$
\langle {\sf O} \rangle \ (= \langle {\sf O}\rangle (s^N) ) = \prod_{i=0}^{N-1} {\sf O} (q^i s). 
$$
As in (\req(avM)) of the generalized $\tau^{(2)}$-model, one can determine the averages of monodromy matrix (\req(mM6d)) of the spin-$\frac{d-1}{2}$ six vertex model as follows:
\begin{prop}\label{prop:6Va} For a positive integer $L$ (no even condition required), 
the average of the $L$th monodromy matrix $(\req(mM6d))$ for $2 \leq d \leq N$ coincides with the classical $L$th monodromy associated to $(\req(s6VL))$; as a consequence, the averages of the entries are given by\footnote{The  $\langle {\sf B} \rangle = 0$ for $d=2$ is the formula (1.36) in \cite{FM01} in the vanishing discussion for the complete $N$-string Bethe ansatz.} 
\be
\langle {\sf A} \rangle  = \langle {\sf D} \rangle = (s^N - s^{-N})^L,  \ \ \langle {\sf B} \rangle = \langle {\sf C} \rangle  = 0. 
\ele(6a)
\end{prop}
{\it Proof.} Parallel to (\req(AL1)) in the generalized $\tau^{(2)}$-model, the following relations about averages of the $(L+1)$- and $L$-th monodromy matrix hold:  
$$
\begin{array}{ll}
\langle {\sf A}_{L+1} \rangle  = \langle {\sf A}_L \rangle \otimes \langle {\sf L}_{0,0}  \rangle + \langle {\sf B}_L \rangle \otimes \langle {\sf L}_{1,0} \rangle + \sum_{k=1}^{N-1} [{\sf A}_L]_{N-k}(s) [{\sf B}_L]_k(q^{N-k} s)\otimes {\sf L}^{(N+1)}_{k, 0}(s) , \\
\langle {\sf B}_{L+1} \rangle  = \langle {\sf A}_L \rangle \otimes \langle {\sf L}_{0,1} \rangle + \langle {\sf B}_L \rangle \otimes \langle {\sf L}_{1,1} \rangle + \sum_{k=1}^{N-1} [{\sf A}_L]_{N-k}(s) [{\sf B}_L]_k(q^{N-k} s)\otimes {\sf L}^{(N+1)}_{k, N}(s) , \\
 {\sf A}_{L+1}, {\sf B}_{L+1}, {\sf A}_L,  {\sf B}_L \longrightarrow {\sf C}_{L+1}, {\sf  D}_{L+1}, {\sf C}_L,  {\sf D}_L ,
\end{array}
$$
where $[{\sf O}]_n (s) := \prod_{i=0}^{n-1} {\sf O}(q^i s)$ for operators ${\sf O}(s)$ and non-negative integers $n$. When $L=1$, the average of monodromy matrix is given by  $\langle {\sf L}_{0,0}  \rangle  = \langle {\sf L}_{1,1}  \rangle  = s^N - s^{-N}$, and $\langle {\sf L}_{0,1}  \rangle  = \langle {\sf L}_{1,0}  \rangle  = 0$. Hence it suffices to show 
${\sf L}^{(N+1)}_{k, 0}(s) = {\sf L}^{(N+1)}_{k, N}(s) = 0 $ for $1 \leq k \leq N-1$.
Since ${\sf L}_{0,0}, {\sf L}_{1,1}$ are interchanged under the conjugation of the spin-inverse operator of $\CZ^d$, and the same for ${\sf L}_{0,1},{\sf L}_{1,0}$, the definition of ${\sf L}^{(N+1)}_{k, 0}$ and ${\sf L}^{(N+1)}_{k, N}$ in turn yield the relation: ${\sf L}^{(N+1)}_{N-k, N}(s)= R^{-1} {\sf L}^{(N+1)}_{k, 0}(s) R $. So one needs only to show ${\sf L}^{(N+1)}_{k, 0} = 0$ for $1 \leq k \leq N-1$. Note that ${\sf L}_{i,j}$'s are $d \times d$ matrices with  ${\sf L}_{i,i}$ diagonal, and ${\sf L}_{i,j} \ (i \neq j)$ upper- or lower-triangular with ${\sf L}_{i,j}^d =0$. This implies 
${\sf L}^{(N+1)}_{k, 0}(s) = 0$ for  $k \geq d$. Hence the first relation in above about the averages of ${\sf A}_{L+1}$ and ${\sf A}_{L}$ for $L=1$ yields
\be
\langle {\sf A}_2 \rangle  = \langle {\sf L}_{0,0} \rangle \otimes \langle {\sf L}_{0,0}  \rangle + \sum_{k=1}^{d-1} [{\sf L}_{0,0}]_{N-k}(s) {\sf L}^k_{0,1} \otimes {\sf L}^{(N+1)}_{k, 0}(s) .
\ele(A2a)
Since ${\sf L}^k_{0,1}$ for $ 1 \leq k \leq d-1$ are linear independent lower-triangular matrices, the invariant of $\langle {\sf A}_2 \rangle$ under $s \mapsto q s$ implies that the same holds for each term in the summation of the above formula. Hence $
[{\sf L}_{0,0}]_{N-k}(s) {\sf L}^k_{0,1} \otimes {\sf L}^{(N+1)}_{k, 0}(s) = [{\sf L}_{0,0}]_{N-k}(qs) {\sf L}^k_{0,1} \otimes {\sf L}^{(N+1)}_{k, 0}(qs)$, then
$$
{\sf L}_{0,0}(s) {\sf L}^k_{0,1} \otimes {\sf L}^{(N+1)}_{k, 0}(s) = {\sf L}_{0,0}(q^{N-k}s) {\sf L}^k_{0,1} \otimes {\sf L}^{(N+1)}_{k, 0}(qs) , \ \ 1 \leq k \leq d-1.
$$ 
Compare the $(k, 0)$th and $(d-1, d-1-k)$th entries of ${\sf L}^k_{0,1}$ in the above equality, then one finds 
$$
a(q^k s) {\sf L}^{(N+1)}_{k, 0}(s) = a(s)  {\sf L}^{(N+1)}_{k, 0}(qs), \ \ a(q^{d-1} s) {\sf L}^{(N+1)}_{k, 0}(s) = a(q^{d-1-k}s) {\sf L}^{(N+1)}_{k, 0}(qs) , 
$$
where $a(s)$ is the function in (\req(6abc)). Hence $
\frac{a(q^k s)}{a(s)} {\sf L}^{(N+1)}_{k, 0}(s) = \frac{a(q^{d-1} s)}{a(q^{d-1-k}s)} {\sf L}^{(N+1)}_{k, 0}(s)$, which by $\frac{a(q^k s)}{a(s)} \neq \frac{a(q^{d-1} s)}{a(q^{d-1-k}s)}$, implies  ${\sf L}^{(N+1)}_{k, 0}(s)=0$ for $1 \leq k \leq d-2$. For $k=d-1$, one has the relation $
a(q^{d-1} s) {\sf L}^{(N+1)}_{d-1, 0}(s) = a(s)  {\sf L}^{(N+1)}_{d-1, 0}(qs)$. Since ${\sf L}^{(N+1)}_{d-1, 0}(s)= \psi (s) {\sf L}^{d-1}_{1,0}$ where $\psi (s)$ is a ${N \choose d-1}$-term sum of $(N-d+1)$-products of $a(q^is)$'s, one has $a(q^{d-1} s) \psi (s) = a(s)\psi (qs)$. Therefore, $\psi (s) =  \alpha \prod_{k=0}^{d-2} a(q^ks) $ where $ \alpha$ is a function invariant under $s \mapsto qs$, hence being a constant scalar  by the degree consideration .
By this, the relation (\req(A2a)) becomes
$$
\langle {\sf A}_2 \rangle  = \langle {\sf L}_{0,0} \rangle \otimes \langle {\sf L}_{0,0}  \rangle + \alpha (\prod_{i=0}^{N-1} a(q^i s)) {\sf L}^{d-1}_{0,1} \otimes {\sf L}^{d-1}_{1,0} , \ \ \alpha \in \CZ.
$$
From the commutation relation of ${\sf A}$ and ${\sf B}$ in (\req(6qdf)), $\langle {\sf A}_2 \rangle $ commutes with ${\sf B}_2(s) \ = ({\sf L}_{0,0} \otimes {\sf L}_{0,1} + {\sf L}_{0,1} \otimes {\sf L}_{1,0})(s) $. Since $[ {\sf B}_2(s), {\sf L}^{d-1}_{0,1} \otimes {\sf L}^{d-1}_{1,0}] \neq 0$, the scalar $\alpha$ is equal to $0$, which implies ${\sf L}^{(N+1)}_{d-1, 0}(s) = 0$. Therefore ${\sf L}^{(N+1)}_{k, 0}(s) = 0$ for all $k$, hence follows the result.
$\Box$ \par  \vspace{.1in}

We now relate the $sl_2$-loop-algebra generators (\req(sl2)) with the vanishing averages of ${\sf B}, {\sf C}$ in (\req(6a)). In later discussions, we shall consider the logarithmic derivative of relations in (\req(6ABCD)).
For simplicity, we shall use the subscripts of variables $s, q, \ldots $ to indicate the logarithmic partial-derivative $s \partial_s , q \partial_q, \ldots$ of operators or functions, e.g, 
${\sf B}_s = s (\partial_s {\sf B})$,  ${\sf B}_q  = q (\partial_q {\sf B})$, $(f_{s, s'})_s = s (\partial_s f_{s, s'})$, $(f_{s, s'})_{s'} = s' (\partial_{s'} f_{s, s'})$ etc. By (\req(6a)), the $s$-derivative of $\langle {\sf B} \rangle, \langle {\sf C} \rangle$ vanishes, and the $q$-derivative of $\langle {\sf B} \rangle , \langle {\sf C} \rangle $  is expressed by
$$
  \langle {\sf B} \rangle_s = 0 , \ \ 
\langle {\sf B} \rangle_q =  \sum_{n=0}^{N-1}{\sf B}_q (sq^n) \prod_{i=0, i \neq n}^{N-1} {\sf B}(sq^i ),  \ ~ \  ~ \  ~ \  {\sf B} \leftrightarrow {\sf C}.
$$
By (\req(Uslq)), the leading and lowest terms of $\langle {\sf B} \rangle_q, \langle {\sf C} \rangle_q$ are given by
\bea(ll)
S^{- (N)} = \frac{\lim_{s \rightarrow 0}    (- s)^{N(L-1)} \langle {\sf B} \rangle_q (s)}{2N(1-\omega) \cdots (1-\omega^{N-1}) } ,& 
T^{- (N)} = \frac{\lim_{s \rightarrow \infty}   s^{-N(L-1)}  \langle {\sf B} \rangle_q (s) }{2N(1-\omega) \cdots (1-\omega^{N-1}) } ; \\
S^{+(N)}  = \frac{ \lim_{s \rightarrow \infty}   s^{-N(L-1)}  \langle {\sf C} \rangle_q (s)}{2N(1-\omega) \cdots (1-\omega^{N-1}) } , &
T^{+ (N)} = \frac{\lim_{s  \rightarrow 0}    (- s)^{N(L-1)} \langle {\sf C} \rangle_q (s)}{2N(1-\omega) \cdots (1-\omega^{N-1}) }  . \\
\elea(SBq)
As in (2.1)-(2.14) of \cite{FM01}, for a given function $\varphi(s)$, there associates the current:
\be
{\sf B}^{(N)} (s) = {\sf B}^{(N)}_1 (s) + {\sf B}^{(N)}_2 (s) 
\ele(BNc)
invariant under $s \mapsto qs$, where ${\sf B}^{(N)}_1 (s)= \langle {\sf B} \rangle_q $, ${\sf B}^{(N)}_2 (s)= \sum_{n=0}^{N-1} \varphi(sq^n) {\sf B}_s (sq^n) \prod_{i=0, i \neq n }^{N-1} {\sf B}(sq^i)$, 
which satisfies the relations: $[{\sf B}^{(N)} (s), B(s')] = [{\sf B}^{(N)} (s), {\sf B}^{(N)} (s')]=0$ (\cite{FM01} (1.39) (1.40)). For a  Bethe state $\Phi (s_1, \ldots , s_m )$, we are going to determine the Fabricius-McCoy current, which means the current (\req(BNc)) for a suitable $\varphi(s)$ with the following property:
\be
{\sf t} (s) (\prod_{l = 1}^{m'} {\sf B}^{(N)} (x_l )) \Phi (s_1, \ldots , s_m ) = (\prod_{l = 1}^{m'} {\sf B}^{(N)} (x_l )) {\sf t} (s)  \Phi (s_1, \ldots , s_m ) 
\ele(Bcur) 
for all $x_l$ and integer $m' \geq 1$. By differentiating relations in (\req(6ADs)), one has
$$
\begin{array}{l}
{\sf A}(s) {\sf B}_{s}(s_n)\prod_{i=1, i \neq n }^m {\sf B}(s_i)
= (\prod_{i=1}^m f_{s, i}) {\sf B}_{s}(s_n)(\prod_{i=1, i \neq n }^m {\sf B}(s_i)) {\sf A}(s) \\
 \ ~ \ ~  - \sum_{k=1 }^m \left(g_{s, k}  \prod_{i=1, i \neq k}^m f_{k, i}\right)_{s_n} {\sf B} (s) ( \prod_{i=1, i \neq k }^m {\sf B}(s_i) ){\sf A}(s_k) 
+ \left(\prod_{i=1}^m f_{s, i} \right)_{s_n} ( \prod_{i=1}^m {\sf B}(s_i) ) {\sf A}(s) \\
 \ ~ \ ~  - \sum_{k=1 }^m g_{s, k} (\prod_{i=1, i \neq k }^m f_{k, i})  {\sf B} (s) \left(\prod_{i=1, i \neq k }^m {\sf B}(s_i) {\sf A}(s_k)\right)_{s_n} ; \\
\left({\sf A}(s) \prod_{i=1}^m {\sf B}(s_i)\right)_q 
= (\prod_{i=1}^m f_{s, i}) \left(\prod_{i=1}^m {\sf B}(s_i) {\sf A}(s)\right)_q - \sum_{k=1}^m \left(g_{s, k} \prod_{i=1, i \neq k }^m f_{k, i}\right)_q {\sf B} (s)  (\prod_{i=1, i \neq k }^m {\sf B}(s_i)) {\sf A}(s_k) \\
 \ ~ \ ~ + \left(\prod_{i=1}^m f_{s, i} \right)_q (\prod_{i=1}^m {\sf B}(s_i)) {\sf A}(s)  - \sum_{k=1}^m g_{s, k} (\prod_{i=1, i \neq k }^m f_{k, i}) ~ \left( {\sf B} (s)  \prod_{i=1, i \neq k }^m {\sf B}(s_i) {\sf A}(s_k) \right)_q ;  \\
  \ ~ \ ~ \ ~ \ ~  \ ~ \ ~   \ ~ \ ~   \ ~ \ ~  {\sf A}(s) , f_{s, i}, f_{k, i}, g_{s, k} \leftrightarrow {\sf D}(s) , f_{i, s},  f_{i, k}, g_{k, s}.
\end{array}
$$
Set $m=N$, $s_i = x q^i $ for $1 \leq i \leq  N (=0) \in \ZZ_N $ in the above relations, then using the relations   
$$
\begin{array}{ll}
f_{k, k+1} = 0 , \ \ \prod_{i=0}^{N-1} f_{s, i} \prod_{i=0}^{N-1} f_{i, s} = 1 , & ( \prod_{i=0, i \neq k }^{N-1} f_{k, i})_q =   ( \prod_{i=0, i \neq k }^{N-1} f_{i, k})_q =   \frac{2 }{q-q^{-1}} ,   \\
( \prod_{i=0, i \neq k}^{N-1} f_{k, i})_{s_n} = ( \delta_{k, n}- \delta_{k, n-1}) \frac{2}{q- q^{-1}}  , & ( \prod_{i=0, i \neq k}^{N-1} f_{i, k})_{s_n} = ( \delta_{k, n}- \delta_{k, n+1}) \frac{-2}{q- q^{-1}} ,
\end{array}
$$
one finds
$$
\begin{array}{l}
{\sf A}(s) {\sf B}^{(N)}_1 (x)
=   {\sf B}^{(N)}_1 (x) {\sf A}(s) 
- \frac{2 }{q- q^{-1}} \sum_{n=0}^{N-1}  g_{s, n} {\sf B} (s) ( \prod_{i=0, i \neq n }^{N-1} {\sf B}(xq^i)) {\sf A}(xq^n ) , \\
{\sf D}(s) {\sf B}^{(N)}_1 (x) = {\sf B}^{(N)}_1 (x) {\sf D}(s) 
+  \frac{2 }{q-q^{-1}} \sum_{n=0}^{N-1}  g_{s, n} {\sf B}(s) ( \prod_{i=0, i \neq n }^{N-1} {\sf B}(xq^i)) {\sf D}(xq^n ) , \\
{\sf A}(s) {\sf B}^{(N)}_2 (x) 
= {\sf B}^{(N)}_2 (x)  {\sf A}(s)  + \frac{2}{q - q^{-1}}  \sum_{n=0}^{N-1} g_{s, n} {\sf B} (s)  (\varphi(xq^{n+1})- \varphi(xq^n) ) ( \prod_{i=0, i \neq n}^{N-1} {\sf B}(xq^i) ){\sf A}(xq^n) , \\
{\sf D}(s) {\sf B}^{(N)}_2 (x)  = {\sf B}^{(N)}_2 (x) {\sf D}(s) 
+ \frac{2}{q- q^{-1}}\sum_{n}  g_{s,n}{\sf B}(s) ( \varphi(xq^{n-1})- \varphi(xq^n))  (\prod_{i=0, i \neq n }^{N-1} {\sf B}(xq^i)) {\sf D}(xq^n) .
\end{array}
$$
Hence 
$$
\begin{array}{l}
{\sf A}(s) {\sf B}^{(N)} (x) 
= {\sf B}^{(N)} (x)  {\sf A}(s)  + \frac{2{\sf B} (s) }{q - q^{-1}}  \sum_{n=0}^{N-1} g_{s, n} (\varphi(xq^{n+1})- \varphi(xq^n)-1 ) ( \prod_{i=0, i \neq n}^{N-1} {\sf B}(xq^i) ){\sf A}(xq^n) ; \\
{\sf D}(s) {\sf B}^{(N)} (x)  = {\sf B}^{(N)} (x) {\sf D}(s) 
+ \frac{2{\sf B} (s) }{q- q^{-1}}\sum_{n=0}^{N-1} g_{s,n} (\varphi(xq^{n-1})- \varphi(xq^n)+1)   (\prod_{i=0, i \neq n }^{N-1} {\sf B}(xq^i)) {\sf D}(xq^n).
\end{array}
$$
The relation $\langle {\sf B} \rangle =0$  yields   
$$
\begin{array}{l}
( \prod_{i=0, i \neq n}^{N-1} {\sf B}(xq^i) ){\sf A}(xq^n) \Phi (s_1, \ldots , s_m )= ( \prod_{i=0, i \neq n}^{N-1} {\sf B}(xq^i) )(a^L(xq^n)\prod_{i=1}^m f_{xq^n, s_i})  \Phi (s_1, \ldots , s_m ) , 
 \\
(\prod_{i=0, i \neq n }^{N-1} {\sf B}(xq^i)) {\sf D}(xq^n)\Phi (s_1, \ldots , s_m ) = 
(\prod_{i=0, i \neq n }^{N-1} {\sf B}(xq^i))(a^L(xq^{n+d-1}) \prod_{i=1}^m f_{s_i, xq^n} ) \Phi (s_1, \ldots , s_m ),
\end{array}
$$
hence the condition (\req(Bcur)) for $m'=1$ is provided by the following constraint of $\varphi(s)$:
\be
(\varphi(sq)- \varphi(s)-1 ) a^L(s) \prod_{i=1}^m f_{s, s_i} + (\varphi(sq^{-1})- \varphi(s)+ 1 ) a^L(sq^{d-1}) \prod_{i=1}^m f_{s_i, xq^n} = 0  
\ele(varpc)
for all $s$. Use the variable $t$ in (\req(6Tjt)) and write the function (\req(p6V)) by $P_{\rm 6V} (t^N) = \sum_{k=0}^{N-1} p(\omega^k t)$ with $p(t) := \frac{H (t)^L t^{-r}  }{F(t)F(\omega  t )}$. Up to the factor $q^{-(d-2)L+m } t^{\frac{-L}{2}+r}F(\omega  t )F( \omega^{-1} t) \prod_{i=0}^{d-3}(1- \omega^i t)^{-L}$, (\req(varpc)) is equivalent to $
(\varphi(sq )- \varphi(s)-1 )   p(\omega^{-1} t)= (\varphi(s)- \varphi(sq^{-1} )- 1 )  p( t )$. 
Hence $\varphi(sq )- \varphi(s)-1 =  \alpha (t^N) p(t)$ for some function $\alpha (t^N)$. The relation $\varphi(sq^N ) = \varphi(s )$ in turn yields $\alpha (t^N) = \frac{-N}{\sum_{i=o}^{N-1} p( \omega^i t) }$, hence the $\varphi$-condition (\req(varpc))   is equivalent to the following equation of $\varphi$:
\be
\varphi(sq )- \varphi(s)-1 = \frac{-N p(t)}{P_{\rm 6V} (t^N) }.
\ele(vapeq)
Up to additive $s^N$-functions, the above equation has the unique solution given by 
\be
\varphi(s) = \frac{- \sum_{k=0}^N k p(\omega t)}{P_{\rm 6V} (t^N)}, \ \ p(t) = \frac{H (t)^L t^{-r}  }{F(t)F(\omega  t )}. 
\ele(solphi)
Note that by $\langle B \rangle_s=0$, all solutions $\varphi(s)$ of (\req(vapeq)) define the same current (\req(BNc)). Therefore we have shown that with $\varphi$ in 
(\req(solphi)), the relation (\req(Bcur)) holds for $m'=1$, then also true for all positive integers $m'$ by the same induction argument  in Sect. 2 D of \cite{FM01}. Hence we obtain the Fabricius-McCoy current given by (\req(BNc)) with $\varphi(s)$ in (\req(solphi)).
Now we show the following result.
\begin{thm}\label{thm:6Vpol} Let $\Phi (s_1, \ldots , s_m )$ be the Bethe state for the solution, $t_i ~ (= q s^2_i), 1 \leq i \leq m,$ of the Bethe equation $(\req(6dBet))$, and $P_{\rm 6V} (\xi)$ be the polynomial $(\req(p6V))$ defined by $t_i$'s. Then the current $E^- ( \xi )$ in $(\req(Ec))$ and the Fabricius-McCoy current ${\sf B}^{(N)} (s)$ in $(\req(Bcur))$ are related by 
$$
(- s)^{N(L-1)} {\sf B}^{(N)} (s) = 2N ( \prod_{i=0}^{N-1}(1-\omega^i)) E^-(t^N) .
$$
As a consequence, $\frac{1}{N} P_{\rm 6V} (\xi)$ is the evaluation polynomial for the $sl_2$-loop-algebra representation determined by $\Phi (s_1, \ldots , s_m )$.
\end{thm}
{\it Proof.}
When acting on the Bethe state $\Phi (s_1, \ldots , s_m )$, the $t^N$-operators, $E^- (t^N)$ and  $(- s)^{N(L-1)} {\sf B}^{(N)} (s)$, both satisfy the property (\req(Bcur)), hence they are proportional to each other. 
By $\langle B \rangle_s=0$, the lowest term of $(- s)^{N(L-1)} {\sf B}^{(N)} (s)$ is contributed only by the term from  ${\sf B}^{(N)}_1 (s)$, which is equal to $2N \prod_{i=0}^{N-1}(1-\omega^i)S^{-(N)}$  by (\req(SBq)); while the current $E^- ( \xi )$ has the constant term 
$E^-(0) = f(0) = S^{-(N)}$. Hence follow the results.
$\Box$ \par \noindent \vspace{.1in}
{\bf Remark.} By the same argument in Sect. III A of \cite{FM01}, the current ${\sf B}^{(N)} (s)$ in the above proposition has poles only at the zeros of $P_{\rm 6V} (t^N)$, which is consistent with that for $E^-(\xi)$.
$\Box$ \par \noindent \vspace{.1in}

\subsection{Comparison of the XXZ chain of spin-$\frac{N-1}{2}$ at $N$th root-of-unity  and superintegrable CPM \label{ssec.6CPM}}
It was noticed \cite{AMP, B93} that the superintegrable CPM and the XXZ chain of spin-$\frac{N-1}{2}$ for $q^N=1$ share the same Bethe equation (up to the phrase factors). In this subsection, we compare the symmetry of the degenerate eigenstates between two models, and provide the answer to a question raised in \cite{NiD}.

Under the change of variables $t = \omega t'$, the polynomial (\req(H6V)) for the XXZ chain of  spin-$\frac{N-1}{2}$ and (\req(HCP)) for superintegrable CPM are related by 
$$
H_{N} ( t ) = H_{\rm CP} (t') , 
$$
by which,  a six-vertex-Bethe-solution $\{ t_i \}_{i=1}^m$  of (\req(6dBet)) is equivalent to a CPM-Bethe-solution  $\{ t'_i \}_{i=1}^m$ of (\req(BeCP)) for sectors, $S^z_{\rm 6V} \equiv  (P_a +P_b)_{\rm CP} = r \pmod{N}$, the  
${\sf T}^{(2)}$-eigenvalue $\lambda^{(2)} (t; t_1, \ldots, t_m)$ in (\req(6lam)) the same as $T^{(2)}(t')$ in (\req(TFfus)), and the identical evaluation polynomials, (\req(p6V)) and (\req(pCP)) : $P_{\rm 6V} (t^N)= \omega^{-r+P_b} {\cal P}_{\rm CP} (t^N)$ (\cite{NiD} Proposition 4.1). In the theory of Onsager algebra symmetry of superintegrable CPM, the polynomial ${\cal P}_{\rm CP} (\xi)$ is well-understood as follows. It is known that
the Onsager algebra can be realized as the Lie-subalgebra of the $sl_2$-loop algebra fixed by a standard $sl_2$-involution and inverting the loop-variable \cite{R91}. 
The (finite-dimensional) Onsager-algebra representation was known in \cite{Dav}, and the rigorous mathematical theory has been fully developed and understood in \cite{DR}, in particular,  the Hermitian irreducible representations have been completely classified. By this,  one finds ${\cal P}_{\rm CP} (\xi )$ is a simple $\xi$-polynomial with only negative real roots and the degree $m_E=$ the integral part $ [\frac{(N-1)L -r-2m}{N}]$, which implies the dimension $2^{m_E}$ for the degenerate $\tau^{(2)}$-eigenspace as an irreducible Onsager-algebra representation (\cite{R05o} Theorem 2 and formula (70)). 
Correspondingly in the XXZ spin chain, one concludes only the spin-$\frac{1}{2}$ representations occur in the $sl_2$-loop-algebra representation generated by the Bethe state $\Phi(s, s_1, \ldots, s_m)$ with $2^{m_E}$-degeneracy for the ${\sf t}^{(2)}$-eigenspace, (by which, follows the answer of the question raised in \cite{NiD} Proposition 4.2). 
From the above discussion, the  root-of-unity XXZ chain of spin-$\frac{N-1}{2}$ and superintegrable CPM  share the equivalent Bethe equation and evaluation polynomial, with the same degeneracy of eigenstates for certain sectors, but they carry the different type of symmetry structure. The Onsager-algebra symmetry of the superintegrable CPM is inherited from the Baxter's $Q$-matrix, indeed generated by the symmetry operators of the quantum
Hamiltonian chain (see \cite{R05o} Sect. 3); while the root-of-unity symmetry of the XXZ spin chain arises from the $q$-derivative (\req(SBq)) of vanishing-average-entries of the monodromy matrix.
It is pertinent to ask whether there exists a larger symmetry algebra than the Onsager algebra in the superintegrable CPM as suggested by the XXZ spin chain. 
But, the answer has not been found yet. The algebraic-Bethe-ansatz discussion of $\tau^{(2)}$-models in section \ref{sec:BABBS} could possibly provide certain clues to this end, though not clear at present.  This is also one of the reasons that we conduct the algebraic-Bethe-ansatz study of $\tau^{(2)}$-model in this work.

\section{Concluding Remarks}\label{sec. F} 
In this article, we have made a systematic account on the algebraic approach about the fusion operators of the generalized $\tau^{(2)}$-model using ABCD-algebra techniques.  The recursive fusion relation (\req(fus)) follows automatically from the construction of fused $L$-operators. 
Under the modest and seemingly innocuous conjecture (\req(LGaug)) supported by computational evidence in cases, we produce a logical explanation about the validity of boundary fusion relation (\ref{fusB}) for the generalized $\tau^{(2)}$-model, by which one can use the separation-of-variables method to solve the eigenvector problem of the model in generic cases ( \cite{GIPS} Theorem 2).  
On two special classes of $\tau^{(2)}$-model centered at the superintegrable one, we perform the algebraic-Bethe-ansatz technique to study the Bethe equation and Bethe states. 
The efforts enable us to reconstruct the $\tau^{(2)}$-eigenvalues of certain sectors among the complete spectra previously known in the theory of superintegrable CPM \cite{AMP, B93} through the functional-relation method. With the similar argument, we discuss the root-of-unity XXZ spin chain of higher spin, and obtain the fusion relations with truncation identity as in the spin-$\frac{1}{2}$ case. 
Furthermore, 
the algebraic-Bethe-ansatz technique can produce better quantitative results in the root-of-unity XXZ spin chain about solutions of  the Bethe equation than those in the $\tau^{(2)}$-model partly due to the symmetric form of the six-vertex $R$-matrix. 
Indeed from the Bethe relation and fusion relations, we successively extract the evaluation polynomial for the $sl_2$-loop-algebra symmetry in the root-of-unity XXZ spin chain of
 higher spin, verified in subsection \ref{ssec.6symd} following the line of the spin-$\frac{1}{2}$ case in \cite{DFM, FM01}. 
Since results in the root-of-unity symmetry of XXZ spin chain bear a remarkable quantitative and semi-qualitative resemblance to the Onsager-algebra symmetry of superintegrable CPM,
it would be desirable to have a Baxter's $Q$-operator for the  XXZ spin chain of higher spin at  roots of unity to pursuit the functional-relation study about the symmetry as in the spin-$\frac{1}{2}$ case \cite{R06Q}. Such $Q$-operator has not been found yet. The aid of a special $Q$-operator in the XXZ spin chain  to encode the  root-of-unity symmetry will provide a useful tool to make a detailed investigation about the degeneracy, much in the same way as the Onsager-algebra symmetry does in the superintegrable CPM. In particular, in the spin-$\frac{N-1}{2}$ case it is possible to construct the Baxter's $Q$-operator of the root-of-unity XXZ spin chain so that the comparison in subsection \ref{ssec.6CPM} will be understood in a more satisfactory manner. A program along this line is under consideration, and progress would be expected. 
In this work, we discuss the symmetry about the generalized $\tau^{(2)}$-model, and the XXZ spin chain  of higher spin. 
We also hope that our results will eventually lead to the understanding of other models, such as the root-of-unity eight-vertex model in \cite{De01a, De01b, FM04, FM41}. This programme is now under progress. For the  root-of-unity symmetry of XXZ spin chain , our discussion can  be applied to a more general setting. But, just to keep things simple, we restrict our attention in this paper only to even $L$, odd $N$ and $S^z \equiv 0 \pmod{N}$, and leave possible generalizations, applications or implications to future work.

\section*{Acknowledgements}
The author is pleased to thank Professor S. Kobayashi for the hospitality in the spring of 2006 during the author's stay at Department of Mathematics, U. C. Berkeley, where part of this work was carried out.
This work is supported in part by National Science Council of Taiwan under Grant No NSC 94-2115-M-001-013.


\begin{thebibliography}{99}
\bibitem{AMP} G. Albertini, B. M. McCoy, and 
J. H. H. Perk, Eigenvalue spectrum of the
superintegrable chiral Potts model, in  Adv. Stud.
Pure Math., 19, Kinokuniya Academic (1989) 1--55.
%
\bibitem{Bax} R. J. Baxter, Exactly solved models
in statistical mechanics, Academic Press (1982).
%
\bibitem{B89} R. J. Baxter, Superintegrable chiral Potts model: Thermodynamic properties, an "Inverse" model, and a simple associated Hamiltonian, J. Stat. Phys. 57 (1989) 1--39.
%
\bibitem{B90} R. J. Baxter, Chiral Potts model: eigenvalues of the transfer matrix, Phys. Lett. A 146 (1990) 110--114.
%
\bibitem{B93} R. J. Baxter, Chiral Potts model with skewed boundary conditions, J.
Stat. Phys. 73 (1993) 461--495.
%
\bibitem{B04} R. J. Baxter, The six and eight-vertex models revisited, J. Stat. Phys. 114 (2004) 43--66; cond-mat/0403138.
%
\bibitem{B049} R. J. Baxter, Transfer matrix functional relation for the generalized $\tau_2(t_q)$ model, J. Stat. Phys. 117 (2004) 1--25;  cond-mat/0409493.
%
\bibitem{B05} R. J. Baxter, Derivation of the order parameter of the chiral Potts model, Phys. Rev. Lett. 94 (2005) 130602; cond-mat/0501227.
%
\bibitem{BazS} V.V. Bazhanov and Yu.G. Stroganov, Chiral
Potts model as a descendant of the six-vertex model, J.
Stat. Phys. 59 (1990) 799--817.
%
\bibitem{BBP} R. J. Baxter, V.V. Bazhanov and
J.H.H. Perk,  Functional relations for transfer
matrices of the chiral Potts model, Int. J. Mod.
Phys. B 4 (1990) 803--870.
%
\bibitem{BIK} N. M. Bogoliubov, A. G. Izegin, and N. A. Kitanine, Correlation functions for a strong correlated boson system, Nucl. Phys. B 516 [FS] (1998) 501--528.
%
\bibitem{ChP} V. Chari and A. Pressley, Quantum affine algebras, Comm. Math. Phys. 142 (1991) 261--283.  
%
\bibitem{DR} E. Date and S. S.  Roan,  The structure of quotients of the 
Onsager algebra by closed ideals, J. Phys. A:
Math. Gen. 33 (2000) 3275--3296,  math.QA/9911018;  The algebraic structure of the Onsager algebra, Czech. J. Phys., Vol.
50 No. 1 (2000) 37--44, cond-mat/0002418.
%
\bibitem{Dav} B. Davies, Onsager's algebra and
superintegrability, J. Phys. A: Math. Gen. 23 (1990)
2245--2261; Onsager's algebra and the 
Dolan-Grady condition in the non-self case, J. Math. Phys.
32 (1991) 2945--2950.
%
\bibitem{DFM} T. Deguchi, K. Fabricius and B. M. McCoy, The $sl_2$ loop algebra symmetry for the six-vertex model at roots of unity, J. Stat. Phys. 102 (2001) 701--736; cond-mat/9912141. 
%
\bibitem{De01a} T. Deguchi: Construction of some missing eigenvectors of the XYZ spin chain at the discrete coupling constants and the exponentially large spectral degeneracy of the transfer matrix, J. Phys. A: Math. Gen. 35 (2002) 879--895; cond-mat/0109078. 
%
\bibitem{De01b} T. Deguchi: The 8V CSOS model and the $sl_2$ loop algebra symmetry of the six-vertex model at roots of unity, Int. J. Mod. Phys. B 16 (2002) 1899--1905; cond-mat/0110121. 
%
\bibitem{De05} T. Deguchi: Regular XXZ Bethe states at roots of unity- as highest weight vectors of the $sl_2$ loop algebra at roots of unity, cond-mat/0503564 v3. 
%
\bibitem{FM01} K. Fabricius and B. M. McCoy, Evaluation parameters and Bethe roots for the six vertex model at roots of unity, {\it Progress in Mathematical Physics} Vol 23, eds. M. Kashiwara and T. Miwa,  Birkh\"{a}user Boston (2002), 119--144; cond-mat/0108057.
%
\bibitem{FM02} K. Fabricius and B. M. McCoy, New developments in the eight vertex model, J. Stat. Phys. 111 (2003) 323--337; cond-mat/0207177.
%
\bibitem{FM04} K. Fabricius and B. M. McCoy, Functional equations and fusion matrices for the eight vertex model, Publ. RIMS, 40 (2004) 905--932; cond-mat/0311122.
%
\bibitem{FM41} K. Fabricius and B. M. McCoy, Root of unity symmetries in the 8 and 6 vertex models, cond-mat/0411419.
%
\bibitem{Fad} L. D. Faddeev, How algebraic Bethe
Ansatz works for integrable models, eds. A.
Connes, K. Gawedzki and J. Zinn-Justin, {\it
Quantum symmetries/ Symmetries quantiques},
Proceedings of the Les Houches summer school,
Session LXIV, Les Houches, France, August 1-
September 8, 1995, North-Holland (1998),  149--219;
%
\bibitem{GIPS} G. von Gehlen, N. Iorgov, S. Pakuliak and V. Shadura: Baxter-Bazhanov- Stroganov model: Separation of variables and Baxter equation, J. Phys. A: Math. Gen. 39 (2006) 7257--7282; nlin.SI/0603028.
%
\bibitem{GR} G. von Gehlen and R. Rittenberg,   
$Z_n$-symmetric quantum chains with infinite set of
conserved charges and $Z_n$ zero modes, Nucl. Phys. B 257
(1985) 351--370.
%
\bibitem{IK} A. G. Izegin. and V. E. Korepin, Lattice model connected with the non-linear Schr\"{o}dinger equation, Sov. Phys. Dokl. 26 (1981) 653--654.
%
\bibitem{Ka} M. Karowski, On the bound state problem in 1+1 dimensional field theories, Nucl. Phys. B 153 (1979) 244--252.
%
\bibitem{KBI} V. E. Korepin, N. M. Bogoliubov, and A. G. Izegin, Quantum inverse scattering method and correlation functions, Cambridge Univ. Press, Cambridge, 1993.
%
\bibitem{Kor} C. Korff: Solving Baxter's TQ-equation via representation theory, in "Non-commutative geometry and representation theory in mathematical physics", Contemporary Math. v391, eds J. Fuchs et al. AMS (2005) 199--211 ; math-ph/0411034. 
%
\bibitem{Krp} I. G. Korepanov, Hidden symmetries in the 6-vertex model of statistical physics, Zap. Nauchn. Sem. S-Petersburg, Otdel. Mat. Inst. Stekelov (POMI) 215 (1994) 163--177; hep-th/9410066.
%
\bibitem{KiR} A, N. Kirillov and N. Yu. Reshetikhin, Exact solution of the integrable XXZ Heisenberg model with arbitrary spin: I. The ground state and the excitation spectrum, J. Phys. A: Math. Gen. 20 (1987) 1565 -- 1595.
%
\bibitem{KRS} P. P. Kulish,  N. Yu. Reshetikhin and E. K. Sklyanin, Yang Baxter equation and representation theory, Lett. Math. Phys. 5 (1981) 393--403.
%
\bibitem{KS} P. P. Kulish and E. K. Sklyanin,
Quantum spectral transform method. Recent
developments, eds. J. Hietarinta and C. Montonen,
Lecture Notes in Physics 151 Springer (1982),
61--119.
%
\bibitem{MR} B. M. McCoy and S. S. Roan, Excitation spectrum and phase structure of the chiral Potts model. Phys. Lett. A 150 (1990) 347--354.
%
\bibitem{Nep} R. I. Nepomechie, Solving the open XXZ spin chain with nondiagonal boundary terms at roots of unity, Nucl. Phys. B622 (2002) 615--632; hep-th/0110116.
%
\bibitem{NiD} A. Nishino and T. Deguchi, The $L(sl_2)$ symmetry of the Bazhanov-Stroganov model associated with the superintegrable chiral Potts model, Phys. Lett. A 356 (2006) 366--370
; cond-mat/0605551.
%
\bibitem{PS} G. P. Pronko and Y. G. Stroganov, Bethe equations 'on the wrong side of the equator', J. Phys. A: Math. Gen. 32 (1999) 2333--2340.
%
\bibitem{R91} S. S. Roan, Onsager's algebra, loop algebra and chiral Potts model, Preprint
Max-Planck-Inst. fur Math.,
Bonn, MPI 91-70, 1991.
%
\bibitem{R04} S. S. Roan, Chiral Potts rapidity curve descended from six-vertex model and symmetry group of rapidities, J. Phys. A: Math. Gen. 38 (2005) 7483--7499; cond-mat/0410011.
%
\bibitem{R05o} S. S. Roan, The Onsager algebra symmetry of $\tau^{(j)}$-matrices in the superintegrable chiral Potts model, J. Stat. Mech. (2005) P09007; cond-mat/0505698.
%
\bibitem{R05b} S. S. Roan, Bethe ansatz and symmetry in superintegrable chiral Potts model and root-of-unity six-vertex model, in Nankai Tracts in Mathematics Vol. 10, {\it Differential Geometry and Physics}, eds. Mo-Lin Go and Weiping Zhang,  World Scientific, Singapore (2006), 399-409; cond-mat/0511543.
%
\bibitem{R06Q} S. S. Roan, The Q-operator for root-of-unity symmetry in six vertex model, J. Phys. A: Math. Gen. 39 (2006) 12303-12325; cond-mat/0602375.
%
\bibitem{Sk} E. K. Sklyanin, Some algebraic structures connected with the Yang-Baxter equation, Funct. Anal. Appl. 16 (1983) 263--270. 
%
\bibitem{TakF} L. A. Takhtadzhan and L. D. Faddeev, The quantum method of the inverse problem and Heisenberg XYZ model, Usp. Mat. Nauk 34 (1979) 13--63 (in Russian), (English Translation: Russ. Math. Surveys 34 (1979) 11--68). 
%
\bibitem{Ta} V. O. Tarasov, Cyclic monodromy matrices for the R-matrix of the six-vertex model and the chiral Potts model with fix spin boundary conditions, Intern. J. Mod. Phys. A7 Suppl. 1B (1992) 963--975.
%
\bibitem{Ta92} V. O. Tarasov, Cyclic monodromy matrices for $sl(n)$ trigonometric R-matrices, Comm. Math. Phys. 158 (1993) 459--483.
\end{thebibliography}
\end{document}